# On the Formation of a Tulip flame in Closed and Semi-open Tubes[1]

Mikhail A. Liberman [a] *, Chengeng Qian[b], Cheng Wang[b]

[a] *Nordita, KTH Royal Institute of Technology and Stockholm University, Hannes Alfvéns väg 12, 114 21 Stockholm, Sweden*
[b] *State Key Laboratory of Explosion Science and Technology, Beijing Institute of Technology, Beijing, 100081, China*

ABSTRACT

The paper examines the mechanism of the tulip flame formation for the flames propagating in closed tubes of various aspect ratios and in a half-open tube. The formation of tulip flames in 2D channels is studied using high resolution direct numerical simulations of the reactive Navier–Stokes equations coupled with a detailed chemical model for a stoichiometric hydrogen/air mixture. The flame propagating in the 3D rectangular channel was studied using Large eddy simulations (LES) and the results were compared with the 2D simulations. In the earlier stages of combustion, the flame accelerates, creating pressure waves and a flow of unburned gas propagating in the same direction as the flame. Subsequently, the lateral parts of the flame surface collapse onto the side walls of the tube, and the flame decelerates, producing rarefaction waves. The interaction of the reverse flow created behind the rarefaction wave front with the unburned gas flow formed earlier by the accelerating flame leads to a decrease in the flow velocity ahead of the flame and an increase in the thickness of the boundary layer. As a result, the axial velocity profile in the near-field flow ahead of the flame takes the form of an inverted tulip with the minimum velocity on the tube axis, which increases towards the side walls, reaching a maximum value and vanishing inside the boundary layer. Since the shape of the flame front is defined by the relative motion of different parts of the flame front, the flame front acquires a tulip shape repeating to a large extent the shape of the velocity profile in the upstream flow. The appearance of rarefaction waves generated by the flame during deceleration stage also explain the flow features observed in experiments, such as reverse flow in unburned and burned gases, reverse pressure gradients, as well as vortex formation and large-scale vortical motion in the burned gas during a tulip flame formation. The whole scenario of the tulip flame formation is qualitatively similar in the 3D and 2D cases but in the 3D case all processes proceed about twice as fast as in the 2D case.

**Keywords:** Tulip flame, flame dynamics, rarefaction wave, pressure wave, vorticity, boundary layer

*Corresponding Author: mliber@nordita.org  (Mikhail Liberman)
Phone: +46855378444 (office)/ +46707692513 (mob)

---

[1] Extended version of the paper #9 presented at the 28th ICDERS, Naples, June 21st, 2022

# 1. Introduction

The inversion of the flame front propagating from the closed end down the tube, from a convex shape with the tip toward the unburned gas to a concave shape with the tip toward the burned gas is known as tulip flame formation and has been observed in many experiments and numerical simulations. The first experimental observation of the tulip-shaped flame formation, and as Dunn-Rankin [1] emphasized, "the most remarkable even to this day photographic study of flame-shape changes in closed vessels" was obtained by O. Ellis [2-4] with the help of then a newly developed camera with rotating shutter. The dynamics of a flame propagating in closed or half-open tubes, is important for understanding combustion processes under confinement, such as explosions and safety issues as well as for industrial and technological applications, e.g. combustion in gas turbines and internal combustion engines.

The name "tulip" flame was given by Salamander et al. in 1957 [5] for a concave flame, which is formed in a tube in experiments resulting in a deflagration to detonation transition (DDT) and this name was subsequently commonly used to describe this particular shape of the flame front. It should be noted that the flame front inversion can be caused by many different processes, such as the interaction of the flame with the shock wave, various hydrodynamic instabilities inherent to the flame, such as Darrieus-Landau (DL) and Rayleigh-Taylor (RT) instabilities etc. This makes the concept of tulip-shaped flame formation not entirely definite, and in particular has led to the involvement of a number of different physical phenomena in various attempts to explain the mechanism of tulip-shaped flame formation since they were first photographed by Ellis [2-4]. The history of earlier attempts to explain the mechanism of the tulip flame formation can be found in the review by Dunn-Rankin [1].

The first attempt to explain the mechanism of tulip flame formation was undertaken by Markstein [6], who suggested that the inversion of the flame shape may be the result of the interaction of a curved flame front with a planar shock wave. Starke and Roth [7] hypothesized that the onset of the tulip flame in Markstein's experiments is associated with the Rayleigh-Taylor instabilities. However, high intensity shock waves, as in Markstein's experiments, do not appear in the early stage of flame propagation, when the tulip shape is formed. From experimental studies of the premixed acetylene/air flame ignited at one end of a closed cylindrical tube, Starke and Roth [7] have stressed that "*First the tulip-shaped structures were found to be coupled with backward motion of the flame and with the sudden decrease in unburned gas velocity in the immediate vicinity of the flame front*."

In an extensive study of tulip flame formation, Guénoche [8] hypothesized that during the deceleration phase "*rarefaction waves are generated, spread in the unburned and the burned gases, and are reflected towards the flame front by the upstream and downstream ends of the tube without changing their sign (i.e. as rarefaction waves); these rarefaction waves give rise to a flow of the unburned gases which may control the behavior of the flame*". Unfortunately, 70 years ago, the level of simulations and experimental methods was insufficient to confirm Guénoche's hypothesis. From the experimental studies of propane/oxygen flame propagation Leyer and Manson [9] also concluded that the onset of vibratory flame propagation is the consequence of the changes in the unreacted gas flow arising from the first contact of the flame front with the side walls of the tube.

More qualitative and quantitative information on the flow velocity, pressure and the flame evolution were obtained in a series of experimental studies by Dunn-Rankin et al. [10-12] and Jeung et al. [13], which led to the hypothesis that the inversion of the flame front may be related to the recirculation of combustion gases observed in the Schlieren images. The quantitative data obtained in experiments [7-13] stimulated a series of numerical studies of the tulip flame phenomenon. Various mechanisms have been proposed in an attempt to explain the inversion of the flame front leading to the tulip-shaped flame. Rotman and Oppenheiem [14] used inviscid numerical simulations considering a potential flow and invoking the corner effect at the back end of the tube as the cause for the vortex structures required for the flame stretching. Marra and Continillo [15] used simulations with zero Mach number and demonstrated that the tulip flame can be formed in the absence of acoustic effects. They also emphasized the influence of the tube aspect ratio on flame propagation, and the effect of wall friction on the onset of a tulip-shaped flame.

Some researchers [1, 9, 16 - 22] suggested that the Darrieus-Landau (DL) instability is the mechanism of the flame front inversion, however this is incompatible with the fact that the characteristic time scale of the DL instability, $\tau_{DL} \cong 1/kU_f$ is much larger than the characteristic time of the flame front inversion observed in experiments and simulations. It should be emphasized that already earlier simulations had shown the presence of a reverse flow in the burned gas behind the flame. In those years, the resolution in numerical simulations was too low to distinguish cause from effect, so simulations could neither confirm nor disprove the involvement of the DL instability in the tulip flame formation. Dold and Joulin [23] demonstrated that numerical solution of the modified Michelson-Sivashinsky equation can show the inversion of the flame front due to the combined effect of the flame front curvature and the DL instability. Matalon and Metzener [24, 25] derived a non-linear evolution equation which describes the motion and instantaneous shape of a two-dimensional flame front as a function of its mean position and contains a memory term related to vortex generation. They demonstrated that numerical solutions to this equation show the inversion of the flame front, which transforms into a tulip-shaped flame. Since the flow in the burned gas is no longer potential, vortices can be created by the curved flame front, and the authors emphasized that it is the vortical motion that can lead to the inversion of the flame front. However, it should be noted that numerical simulations with zero viscosity by Dunn-Rankin et al. [11] and Rotman and Oppenheiem [14] demonstrated the formation of a tulip flame without vorticity.

Clanet and Searby [26] conducted extensive experimental and analytical studies of flames in a half-open tube. They showed that the time of tulip flame formation depends mainly on the laminar flame velocity, the tube radius and aspect ratio L/D of the tube, indicating that the structure of the boundary layer may affect the process. They also distinguished four stages in the flame evolution: the ignition of a spherical flame, the finger-like flame, the quenching of the lateral flame skirt at the side walls causing a decrease in flame speed, and the tulip flame formation stage, and suggested that the formation of the tulip flame is a manifestation of Rayleigh-Taylor instability.

In recent years, interest in flame dynamics in closed and semi-open tubes has increased dramatically due to the need for a better understanding of combustion in engines, as well as flame dynamics during the transition from deflagration to detonation, which is required to solve safety problems. As a result, numerous experimental and numerical studies of the formation of a tulip-shaped flame have been published. However, despite great efforts, there is still no known physical mechanism that causes the formation of a tulip flame, which would be convincingly confirmed by numerical simulation or experiment. Of particular note are the experimental studies of Ponizy et al. [27], from which it follows that the formation of a tulip flame is a purely hydrodynamic phenomenon, in which the intrinsic flame front instabilities are not involved in the process.

Hariharan and Wichman [28, 29] explained the formation of the tulip flame as following: "*The factor having greatest importance for altering the flame structure and behavior is the movement of the singular (saddle) points. The formation and movement of these singular points is dictated by the presence of vorticity in the channel*" However, this is not a physical process which causes the flame front inversion but a simple description of the observed features of the flow formed during the tulip-shaped flame formation. Xiao et al. [30] came to conclusion that "*the circulation near the flame front in the burnt gas is the dominant physical process involved *in the formation of the distorted tulip flame.*" Another publication by Xiao et al. [31], although devoted to numerical studies of the distorted tulip flame, but was also focused on the formation of the tulip flame preceding the distorted tulip flame. The conclusion of [31] was that "*our simulations support the mechanism of vortex motion described by Matalon and Metzener. Large-scale vortices appear just behind the flame skirt near the tube sidewalls after the flame skirt touches the sidewalls. The vortices expand with time and overtake the flame front. Reverse flow forms and dominates the near-field region due to the vortex motion. This process creates conditions that allow the flame to invert*". We emphasize that the flow features observed in simulations [28–31], such as a saddle point, large-scale vortex motion, the adverse pressure gradient, etc., are a consequence of physical processes that cause the formation of a tulip flame, but these features by themselves do not cause the formation of a tulip flame. It is worth noting also that parameters of the hydrogen-air flame used in simulations [31]: the laminar flame speed $U_f = 3.11 \, \text{m/s}$ and adiabatic constant $\gamma = 1.17$ differ significantly from the experimental values $U_f = 2.6 \, \text{m/s}$ and $\gamma = 1.39$.

In the present paper we present the results of 2D numerical simulations (DNS) of a premixed stoichiometric hydrogen/air flame ignited at the closed end of the tube and propagating in tubes of various aspect ratios with both ends closed and in half-open tubes, as well as the results of 3D Large Eddy Simulations (LES). The 2D simulations used a high-order numerical method to solve the fully compressible reactive Navier–Stokes equations coupled with a detailed chemical model for $H_2$/air. The results obtained show that the key physical process for the formation of a tulip flame is the rarefaction wave generated by the flame in unburned gas at the deceleration stage. The interaction of the reverse flow created behind the rarefaction wave front with the unburned gas flow formed earlier by the flame at the acceleration stage leads to a decrease in the flow velocity in front of the flame and an increase in the thickness of the boundary layer. As a result, the axial velocity profile in the near-field flow ahead of the flame takes the form of an inverted tulip with a minimum velocity on the pipe axis, which increases towards the side walls, reaching a maximum value and vanishing inside the layer boundary. Since the shape of the flame front is determined by the relative motion of the various parts of the flame front, the flame front takes on a tulip shape, largely following the shape of the axial velocity profile in the upstream flow created by the rarefaction wave. It is also worth noting that the flow features observed in simulations, such as reverse flow in unburned and burnt gases, reverse pressure gradients, and large-scale vortex motion in burnt gas during the formation of a tulip flame, are a consequence of the rarefaction waves generated by decelerating flame.

## 2. Evolution of a flame propagating in a channel with wall friction

The flame ignited on the center line of a tube near the closed end and propagating to the opposite closed or open end, acquires a convex shape with the tip of the flame front pointing towards the unburned gas. As mentioned earlier, Clanet and Searby [26] distinguished four stages in the flame evolution: the spherical flame, the finger-like flame, the quenching of the lateral flame skirt at the side walls, and the tulip flame formation stage. Later, following the Clanet and Searby work, further analytical development of the 2D theoretical model [26] were made by Bychkov et al. [32, 33] assuming a parabolic velocity profile in the unburned flow. Although an analytical model does not take in consideration all the details of the flow, it sheds light on the features of the dynamics of the flame propagating in tubes with no-slip walls, which are essential for understanding the combustion regimes under confinement.

From the very beginning thermal expansion of the high temperature combustion products pushes the unburned gas towards the opposite end of the tube creating a flow of the unburned gas ahead of the flame, which leads to the flame acceleration compatible with the boundary conditions [34]. After the ignition of the flame near the closed end of the tube, the flame front expands rapidly, taking on an almost hemispherical (semi-circular in the 2D case) shape. At the ignition stage the flame front expends with the velocity $(\Theta-1)U_f$ and after a short time $t_{sph} \sim D/2(\Theta-1)U_f$, the next stage of a finger-shaped flame begins, when the flame accelerates exponentially due to the increase of the lateral surface of the flame skirt

$$X_{tip} \propto \exp(2\Theta U_f t / D), \qquad (1)$$

where $D$ is the channel width, $\Theta = \rho_u / \rho_b$ is the ratio of the densities of unburned $\rho_u$ and burned, $\rho_b$ gases, and $U_f$ is the laminar flame speed.

Depending on the size (radius) of the originally ignited spherical flame, the laminar flame speed, and the width of the tube, the trailing edges of the flame skirt may touch the side walls of the tube before the finger flame begins. Expansion of the high temperature burnt gas behind the flame front causes a flow of unburned gas moving in the same direction as the flame. Therefore, the velocity of every part of the flame front in the laboratory reference system is the sum of the laminar flame velocity relative to the unreacted mixture and the local velocity of the unburned gas flow immediately ahead of this part of the flame front, from which this part of the flame front is drifted by the unburned gas flow.

$$U_{fl} = U_f + U_+(x, y). \qquad (2)$$

In a wide tube the velocity of the flow ahead of the flame is almost constant in the bulk cross section of the tube and drops to zero at the side walls inside the boundary layer of thickness $\delta_l \approx 5X/\sqrt{\mathrm{Re}}$, where $\mathrm{Re} = U_+ X/\nu$ is the Reynolds number, $X$ is the coordinate along the tube, $\nu$ is kinematic viscosity. The shape of the flame front is defined by the relative motion of different parts of the flame front. As the flame front advances into a nonuniform velocity field, the flame surface is stretched tending to reproduce the shape of the velocity profile in the flow ahead of the flame, and the flame surface increases. The stretched flame consumes fresh fuel over a larger surface area which results in an increase in the rate of heat release per frontal area of the curved (wrinkled) flame sheet. The increase in the rate of heat release due to the flame stretching gives rise to a higher volumetric burning rate, and a higher effective burning rate (combustion wave velocity) based on the average heat release rate per frontal area of the stretched flame sheet. A higher burning rate results in an enhancement of the flow velocity ahead of the flame, which in turn gives rise to a larger gradient field and enhanced flame stretching, and so on. In this way a positive feedback coupling is established between the flow of unburned gas and the burning velocity as the flame is stretched in the velocity field ahead of the flame front. The shape of the flame front repeats the shape of the velocity profile of the unburned gas flow, remaining practically flat in volume (in fact, retaining the convex shape obtained during ignition), with the trailing edges of the flame skirt stretched back in the boundary layer. In the framework of a thin flame front model, the increase of the burning rate is proportional to the relative increase of the flame surface area (length in 2D case), which grows linearly in time with accuracy $\delta_l / D << 1$ and which in turn gives rise to a larger gradient of the velocity field. Therefore, the combustion wave velocity increases exponentially in time as [35]

$$U_{fl} \propto \exp(\alpha \Theta U_f t / D) \qquad (3)$$

where $\alpha$ is a numerical factor of the order of unity. It should be noted that expressions (2) and (3) are almost identical, therefore, both in experiments and in numerical simulations, it is rather difficult to distinguish whether the acceleration stage corresponds to a finger-shaped flame or stretching of the lateral flame skirt along the boundary layer.

The lateral part of the flame skirt stretching backwards along the tube wall forms a fold with the side wall of the tube. As the flame speed increases, the flow velocity in the unburned gas also increases and the thickness of the boundary layer decreases. When the thickness of the boundary layer becomes of the order of the flame front thickness $L_f$, which is the characteristic scale of thermal conduction, the temperature of the unburned gas inside the fold increases significantly, and the unburned fuel inside the fold will be burned out very fast. This leads to the collapse of the lateral part of the flame skirt and a sharp decrease in the flame speed. In a sense the collapse of the fold resembles a cumulative jet and looks like a small explosion that generates weak pressure waves diverging from the side wall. The effect of these pressure waves is small compared to the rarefaction waves produced by the decelerating flame. In the case of isothermal boundary conditions, the scenario is similar: when the boundary layer thickness becomes about $L_f$, the lateral part of the flame skirt is quenched at the tube wall.

The flame front can be thought of as a piston that generates pressure waves during acceleration and rarefaction waves during deceleration. The rarefaction waves generated by decelerating flame propagate forward (in the positive direction) in the unburned gas. The front of the rarefaction wave propagates in the positive direction with the speed of sound, while the flow velocity created by the rarefaction wave is in the opposite (negative) direction. Therefore, effect of the rarefaction wave is a decrease of the flow velocity in the unburned flow or the formation of the reverse flow in the unburned gas, depending on the intensity of the wave. The decrease of the flow velocity, especially in the near-field ahead of the flame front, leads to the increase of the boundary layer's thickness, so that the axial velocity profile in the unburned gas takes a shape of an inverted tulip. This leads to a maximum decrease in the flame front velocity at the center line, where the axial velocity in the near-field flow of the unburned gas is minimal, with a gradual increase towards the side walls of the tube and vanishing inside a wide boundary layer. This scenario means that the tips of the flame tulip petals will be located on the inner edges of the boundary layer. In the following sections, we will show that numerical simulations fully confirm the described mechanism for the formation of a tulip-shaped flame.

Dunn-Rankin [1] noticed that for a tulip flame to form, the width of the tube should be large enough (>25 mm) to obviate the wall-heat transfer and boundary-layer effects, but small enough (<100 mm) to prevent large buoyancy effects. While the upper limit is correct, there is another and stronger limitation for the smallest width of the tube.

The difference in the dynamics of a flame propagating propagation in a "wide" and in a "narrow" tube is due to the different time required for establishing a Poiseuille flow with a parabolic velocity profile in the unburned gas ahead of the flame. The width of a channel for which the parabolic velocity profile is formed can be estimated from the condition that the thickness of the boundary layer $\delta_l \approx 5X/\sqrt{Re}$ will become of the order of a half width of the channel. For estimates it is convenient to express the Reynolds number in terms of the flame thickness and the laminar flame velocity [35, 36], then, taking $L_f U_f \approx \nu$ we obtain $Re \approx \Theta D/L_f \approx 25$. The condition for the Poiseuille flow formation then will be $D \approx 25 L_f / \Theta$ which for hydrogen/air flame ($L_f = 0.35 mm$, $\Theta = 7.8$, $U_f = 2.36 m/s$) gives $D \approx 25 L_f / \Theta = 1.1 mm$. These estimates are in a good agreement with experimental study of DDT in microscale tubes [37, 38], which shows that in narrowed tubes (D < 1mm) the flame velocity increases exponentially up to a transition to detonation, bypassing the deceleration stage.

The characteristic time for establishing a Poiseuille flow ahead of the flame can be estimated as $t_P \approx D^2/100\nu$ [35, 36, 39]. For a hydrogen-air ($\nu = 0.22 cm^2/s$) a parabolic velocity profile in the flow ahead of the flame in a "wide" tube $D = 1 cm$ establishes in time $t_P \approx 50 ms$, which is much longer than the time of the tulip flame formation (1-2)ms. On the other hand, in a narrow tube $D = 1 mm$, the parabolic velocity profile is established in $t_P \approx 0.5 ms$. That's why the tulip flame has never been observed in narrow tubes.

## 3. Numerical models
### 3.1 Two-dimensional DNS modeling

We consider a laminar stoichiometric hydrogen-air flames ignited near the left closed end of the two-dimensional rectangular channel and propagating towards the opposite closed or open end. The numerical simulations solve the 2D time-dependent, reactive compressible

Navier-Stokes equations including molecular diffusion, thermal conduction, viscosity and detailed chemical kinetics. The governing equations are

$$\frac{\partial \rho}{\partial t} + \frac{\partial (\rho u_i)}{\partial x_i} = 0, \tag{4}$$

$$\frac{\partial (\rho u_i)}{\partial t} + \frac{\partial (P\delta_{ij} + \rho u_i^2)}{\partial x_j} = \frac{\partial \tau_{ij}}{\partial x_j}, \tag{5}$$

$$\frac{\partial (\rho E)}{\partial t} + \frac{\partial [(\rho E + P)u_i]}{\partial x_i} = \frac{\partial (\tau_{ij} u_i)}{\partial x_i} - \frac{\partial q_i}{\partial x_i}, \tag{6}$$

$$\frac{\partial \rho Y_k}{\partial t} + \frac{\partial \rho u_i Y_k}{\partial x_i} = \frac{\partial}{\partial x_i}(\rho V_{ik} Y_k) + \dot{\omega}_k, \tag{7}$$

$$P = \rho R_B T \left( \sum_{i=1}^{N_s} \frac{Y_i}{W_i} \right), \tag{8}$$

Here $\rho$, $u_i$, $T$, $P$, $E$, $\tau_{ij}$, $q_i$ are density, components of velocity, temperature, pressure, specific total energy, viscosity stress, heat flux mass, $R_B$ is the universal gas constant. $Y_i, W_i, V_{i,j}$ are the mass fraction, molar mass and diffusion velocity of species $i$. The viscosity and thermal conductivity of pure-species and mixture are calculated by the Chapman-Enskog expression [40] and a semi-empirical formula [41].

The reaction rate of species $i$ is determined as

$$\dot{\omega}_k = W_i \sum_{j=1}^{N_r} (v''_{jk} - v'_{jk}) \cdot \left( k_{f,j} \prod_{k=1}^{N_s} \left( \frac{\rho Y_k}{W_k} \right)^{v'_{jk}} - k_{b,j} \prod_{k=1}^{N_s} \left( \frac{\rho Y_k}{W_k} \right)^{v''_{jk}} \right) \tag{9}$$

where $v'_{jk}$ and $v''_{jk}$ are stoichiometric coefficients of species $k$ of the reactant and product sides of reaction $j$. A detailed chemical kinetic model for hydrogen/air developed by Kéromnès et al. [42] was implemented in the simulations. This chemical model has been extensively tested against experimental data and found to accurately predict ignition delay and laminar flame velocity over a wide range of pressures (1–70 bar), temperatures (900–2500 K), and equivalence ratio. (0.1-4.0).

The initial conditions are $P_0 = 1\,atm$, $T_0 = 298\,K$. It should be emphasized that combustion in closed tubes is accompanied by a pressure buildup. During a tulip flame formation pressure increases from initial value $P_0 = 1\,atm$ up to ~1.5 atm for a high aspect ratio and up to ~2.0 atm in a short tube.

### 3.2 Three-dimensional LES modeling

The computational cost of three-dimensional simulations resolving properly all characteristic scales is currently very high. An efficient approach is to use Large Eddy Simulation (LES). In LES, all quantities are filtered in the spectral space by applying filter to the governing equations. Since the flow is compressible, the mass-weighted Favre is introduced. The filtered governing equations are

$$\frac{\partial \bar{\rho}}{\partial t} + \frac{\partial \bar{\rho} \tilde{u}_i}{\partial x_i} = 0, \tag{10}$$

$$\frac{\partial \bar{\rho} \tilde{u}_i}{\partial t} + \frac{\partial}{\partial x_j}\left(\bar{\rho}\tilde{u}_i\tilde{u}_j + \bar{P}\delta_{ij}\right) = \frac{\partial \bar{\tau}_{ij}}{\partial x_j} - \frac{\partial \tau_{ij}^{sgs}}{\partial x_j}, \tag{11}$$

$$\frac{\partial \bar{\rho}\tilde{E}}{\partial t} + \frac{\partial}{\partial x_i}(\bar{\rho}\tilde{E}+\bar{P})\tilde{u}_i = \frac{\partial(\tilde{u}_i \bar{\tau}_{ij})}{\partial x_i} - \frac{\partial \bar{q}_i}{\partial x_i} - \frac{\partial}{\partial x_i}H_i^{sgs} + \frac{\partial}{\partial x_i}\sigma_{ij}^{sgs}, \tag{12}$$

$$\frac{\partial \bar{\rho}\tilde{Y}_k}{\partial t} + \frac{\partial}{\partial x_i}(\bar{\rho}\tilde{Y}_k\tilde{u}_i) = \frac{\partial}{\partial x_i}\left(\bar{\rho}\tilde{D}_i \frac{\partial \tilde{Y}_k}{\partial x_i}\right) - \frac{\partial \varphi_i^{sgs}}{\partial x_i} + \bar{\omega}_k \tag{13}$$

The unresolved fluxes $\sigma_{ij}^{sgs}$ are negligible, and the unresolved Reynolds stresses $\tau_{ij}^{sgs}$ are modeled according to the Boussineq assumption

$$\tau_{ij}^{sgs} - \frac{1}{3}\delta_{ij}\tau_{kk}^{sgs} = -\rho v_t \left(\frac{\partial \tilde{u}_i}{\partial x_j} + \frac{\partial \tilde{u}_j}{\partial x_i} - \frac{2}{3}\delta_{ij}\frac{\partial \tilde{u}_k}{\partial x_k}\right). \tag{14}$$

The sub-grid viscosity $v_t$ is calculated by wall-adapting local eddy viscosity model and the sub-grid heat flux is defined by [43]

$$H_i^{sgs} = -\bar{\rho}\frac{v_t c_p}{\text{Pr}_t}\frac{\partial \tilde{T}}{\partial x_i}, \qquad (15)$$

where $\text{Pr}_t = 0.75$ sub-grid Prandtl number.

In simulations of turbulent reactive flow, the most important is the combustion model, which is used to approximate the filtered reaction rates. In this paper, we employ the so-called thickened flame model [44]. The flame front is thickened artificially without changing the laminar flame velocity. For a premixed laminar flame, the laminar flame velocity $U_f$ and the flame thickness $L_f$ can be expressed as

$$U_f \propto \sqrt{D_{th} A}, \quad L_f \propto \frac{D_{th}}{U_f} = \sqrt{D_{th}/A}, \qquad (16)$$

where $D_{th}$ and $A$ are thermal diffusivity and pre-exponential factor. Supposing that $D_{th}$ and $A$ are replaced by $FD_{th}$ and $A/F$, it is easy to check that $L_f$ increases by a factor $F$ while $U_f$ remains unchanged. A wrinkling factor function $\varepsilon$ is introduced to model the flame front wrinkled by sub-grid scale vortexes. A power-law wrinkling factor function derived by Charlette [45] is employed in simulations. Appling the thickened flame model, the filtered Eq. (13) reads

$$\frac{\partial \bar{\rho}\tilde{Y}}{\partial t} + \frac{\partial}{\partial x_i}(\bar{\rho}\tilde{Y}\tilde{u}_i) = \frac{\partial}{\partial x_i}\left(\bar{\rho}\varepsilon\tilde{D}_i\frac{\partial \tilde{Y}}{\partial x_i}\right) + \frac{\varepsilon\bar{\omega}}{F} \qquad (17)$$

In both 2D and 3D simulations we used adiabatic no-slip reflecting boundary conditions at the walls of the tube:

$$\vec{u} = 0, \ \partial T/\partial \vec{n} = \partial Y_k/\partial \vec{n} = 0, \qquad (18)$$

where $\vec{n}$ is the normal to the wall.

### 3.3 Numerical schemes and modeling parameters

The 2D direct numerical simulations were performed using the DNS solver, which used the fifth order weighted essentially non-oscillatory (WENO) finite difference schemes [46] to resolve the convection terms of the governing equations. The advantage of the WENO finite difference method is the capability to achieve arbitrarily high order accuracy in smooth regions while capturing sharp discontinuity. To ensure the conservation of the numerical solutions, the fourth order conservative central difference scheme is used to discretize the non-linear diffusion terms [47]. The time integration is third order strong stability preserving Runge–Kutta method [48]. The resolution and convergence (a grid independence) tests similar to that in our previous publications [49, 50] were thoroughly performed to ensure that the resolution is adequate to capture details of the problem in question and to avoid computational artefacts.

The reliable simulations of reactive flows require a proper resolution of the inner structure of a flame. According to [49, 50], a resolution of 8 grids per flame front is sufficient to predict the correct laminar flame velocity. Since the thickness of the laminar flame decreases with increasing pressure, higher resolution is needed to resolve the flame structure at maximum pressure. In our simulations, the maximum pressure during the tulip flame formation is less than 2 bar, which leads to $L_f \simeq 130 \mu m$. In the two-dimensional simulations we used resolution the $\Delta x \simeq 12.5 \mu m$, which corresponds to 30 grid points over the flame width at the beginning of the process, and over 15 grid points at the maximum pressure. In the three-dimensional large eddy simulations, we used a coarse mesh of width $\Delta x \simeq 125 \mu m$ defined as a function of pressure which ensures that the flame front is well resolved at given pressure.

It should be noted that the main features of the flame dynamics observed experimentally during the formation of a tulip flame were also observed, at least qualitatively, in many 2D simulations. However, as mentioned earlier in the Introduction, these observations have been limited to superficial descriptions in general terms, without analysis of the physical processes that lead to and explain the flame front flattering, followed by its inversion and the formation of a tulip flame, and generally to incorrect conclusions about the mechanism of tulip-shaped flame formation.

## 4. Results of simulations

Two-dimensional and three-dimensional numerical simulations, were performed to find out and clarify physical processes, which lead to the flattering and inversion of the flame front and the formation of a tulip flame, for four computational domains of rectangular channels of the width $D = 1\,\text{cm}$ with both closed ends for various aspect ratios $L/D = 6, 12, 18$ and for the half-open channel with the right end open. The propagating flame was ignited at the center of the left closed end of the tube by a small half-circular flame in the 2D case and by a hemispherical flame in the 3D case with a hot, burnt gas behind the flame front. The no-slip adiabatic boundary conditions were used at the tube walls.

### 4.1 Tulip flame formation in a short tube: L/D = 6

Figure 1 shows a sequence of numerical images of Schlieren patterns with the stream lines for selected times, which show the evolution and dynamics of the flame and flows in the channel $D = 1\,\text{cm}$, $L/D = 6$ during the development of a tulip flame. It can be seen that the reverse flow in the burned gas is formed immediately after the beginning of the flame deceleration, which is caused by the rarefaction wave generated by the decelerating flame in the unburned gas. An important consequence is that the flow velocity in the near-field ahead of the flame front and respectively the local velocity of the flame front decreases stronger at the center line and lesser toward the wall up to the inner edge of the boundary layer, at $y = 0.4\,\text{cm}$. It should be emphasized that the rarefaction wave changes the structure of the flow in the near-field ahead of the flame front: the considerable increase in the thickness of the boundary layer and the velocity profile. The reverse flow is formed in the near-field region of the unburnt gas and the difference between velocities in the unburnt flow $U_+(x, y = 0.4)$ and $U_+(x, y = 0)$ increases. As a result, the closer to the central line, the stronger the local velocity of the flame front decreases compared to the velocity of the flame front at $y = 0.4 \text{cm}$.

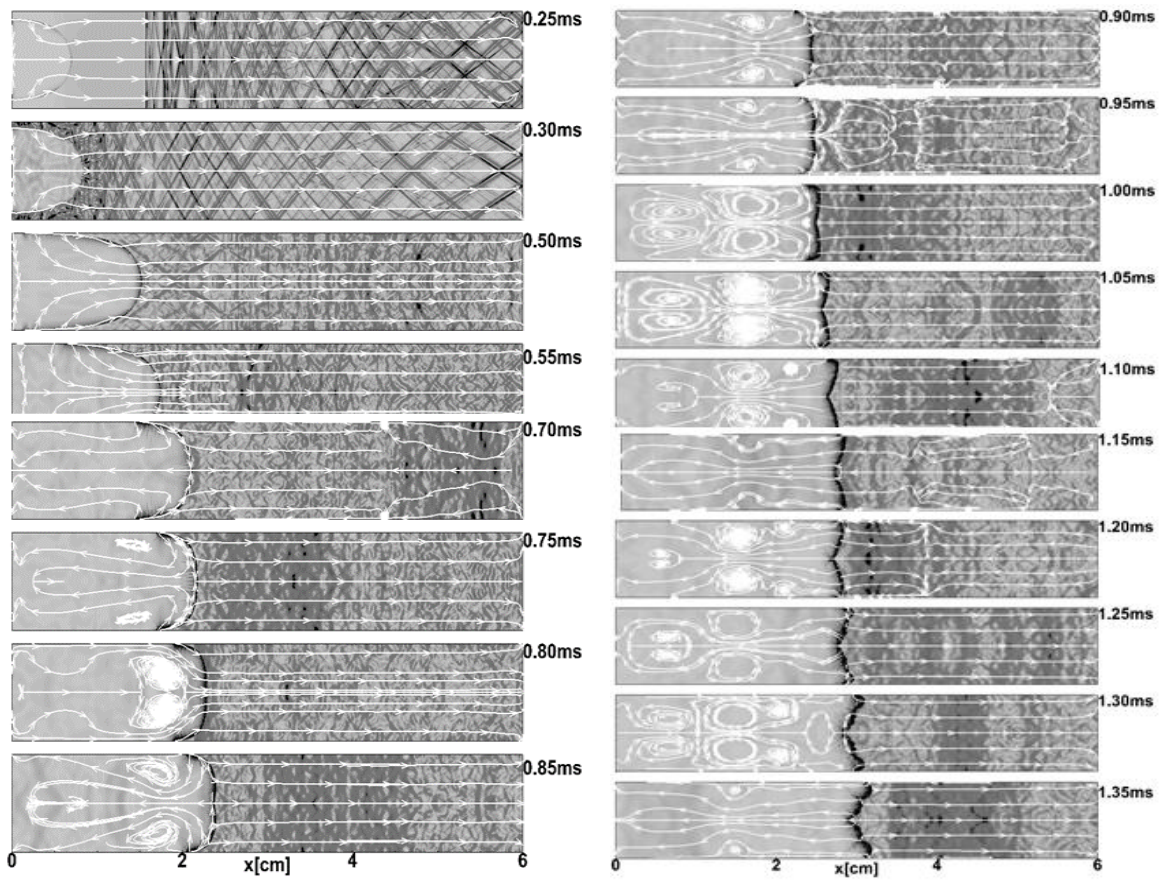

**Fig. 1**(a,b-c,d). Time sequence of computed Schlieren images and streamlines for the flame propagating in the tube $L/D = 6$ with both ends closed.

It is seen that pressure waves produced by the flame during acceleration stage, reflect from the sidewalls, form the ends of the tube and propagate back and forth along the tube.

Figure 2 shows the time evolution of local velocities on the flame front at the center line $U_{fl}(y = 0)$ and near the side wall $U_{fl}(y = 0.4\text{cm})$, the combustion wave velocity (burning rate), $S_f$, and the flame surface area $F_f$. Up to 0.6ms, the combustion wave velocity (burning rate) and the flame tip velocity (the point on the flame front at the center line) increase more or less synchronously and reach their maximum values at 0.55ms. The lateral parts the flame skirt begins to collapse at the side wall at 0.5ms, and the deceleration of the combustion wave and the flame tip begins 0.05ms later. The deceleration rate increases with a significant decrease in the flame surface area at 0.6 ms. The change in the flame acceleration rate at (0.3-0.4) ms is due to the collision of the flame with pressure waves, reflected from the right end of the tube. In a longer tube the collision of the flame with pressure waves occurs later and has a weaker effect. This slows down the flame and explains higher rate of the flame deceleration compared with tubes with higher aspect ratio. After 0.6ms the flame surface area, the combustion wave velocity and velocity of the flame tip decrease, while the local speed of the flame front near the wall $U_{fl}(y=0.4 \text{cm})$ increases and at 0.90ms exceeds $U_{fl}(y=0)$.

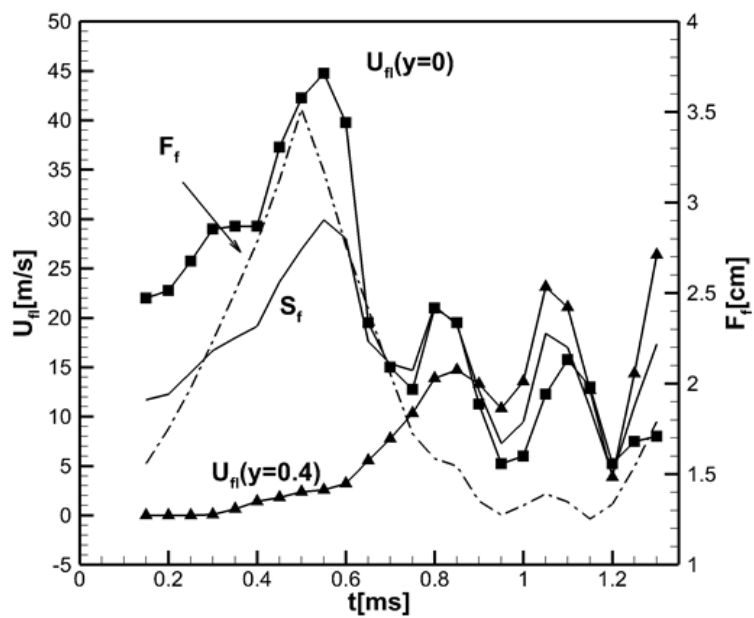

**Fig. 2**. Time evolution of combustion wave velocity $S_f$, local velocities of the flame front at $y=0$ and at $y=0.4 cm$ and the flame surface area $F_f$ (dashed-dotted), tube $L/D=6$.

The flame propagation speed in the laboratory reference frame is closely related to the variation of the flame surface area, as it is seen in Figs. 2. It is seen in Schlieren images (Fig. 1) that oblique pressure waves generated by a convex flame front at the stage of flame acceleration, propagate through the tube, repeatedly colliding and reflecting from the sidewalls, reflect from the right end of the tube and run back and forth along the tube. This leads to small fluctuations in pressure and flame speed.

Figures 3(a, b) show the time evolution of the local velocities of the flame front along the center line $U_{fl}(y=0)$ and near the boundary layer $U_{fl}(y=0.4 \text{cm})$, and the flow velocity in the unburnt gas at 1mm and 3mm ahead of the flame front at $y = 0$ and at $y = 0.4 \text{cm}$. It is seen that velocity of the flow ahead of the flame and the local velocity of the flame front become greater than the flow velocity and the flame front velocity near the center line. As a result, the central part of the flame front will begin to lag behind compared to the lateral parts of the flame.

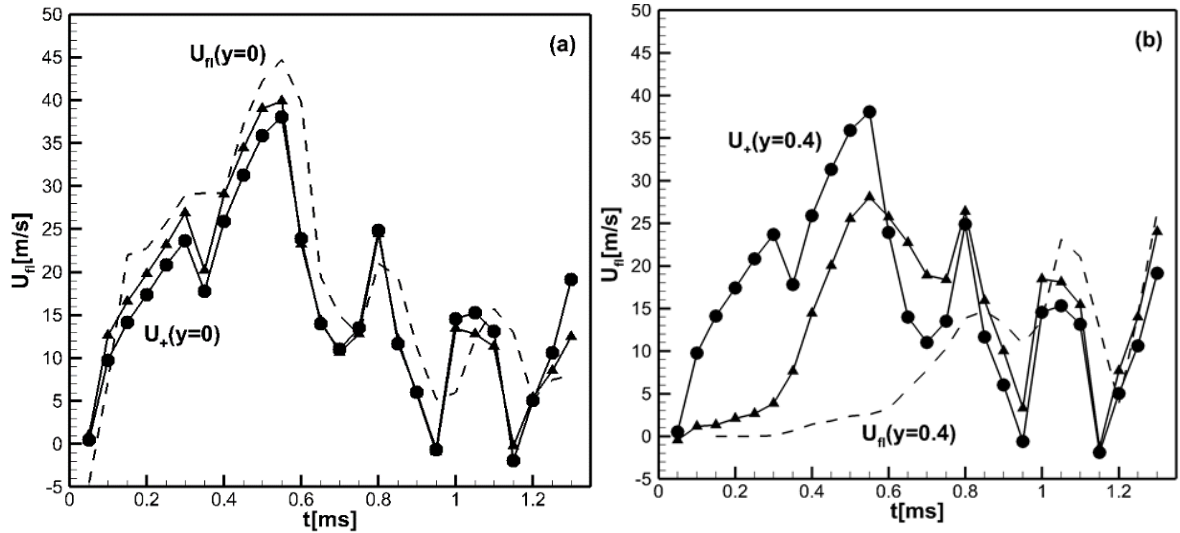

**Fig. 3 (a, b).** a) Flame front velocity at $y = 0$ (dashed line) and flow velocity at 1mm (▲) and 3 mm (●) ahead of the flame front. b) Flame front velocity at $y = 0.4 \text{cm}$ (dashed line) and flow velocity 1mm (▲) and 3 mm (●) ahead of the flame front.

Note that a decelerating flame generates a rarefaction wave in the unburnt gas, the front of which propagates to the right end of the tube at the speed of sound, with the flow velocity behind the front in the opposite (towards the flame) direction. Contrary to the accelerating flame which creates negative pressure gradients supporting the unburnt flow ahead of the flame, the pressure gradient in the refraction wave is positive [2]. During the development of the tulip flame the velocities and acceleration in the various parts of unburnt flow and the flame front change from positive to negative values. This leads to the formation in front of the flame a complex unsteady flow with a reverse flow, which is formed and dominates in the near-field zone ahead of the flame front with the maximum value of the axial velocity $U_+$ on the tube axis, $y=0$.

---

[2] Formally, this is true for a piston that starts moving with negative acceleration in a gas at rest (a piston that begins to move with negative acceleration out from a tube filled with gas at rest [51]). However, this is correct at least qualitatively in our case also.

The difference between the flow velocities at 1 mm ahead of the various points of the flame front at the cross section of the tube: $y=0;\ 0.25\,\text{cm};\ 0.4\,\text{cm}$, is shown in Figure 4(a,b). The change, for example, $\Delta U_+ = U_+(y=0.4) - U_+(y=0)$ is the consequence of the flame front acceleration or deceleration and corresponds to the stages when the accelerating or decelerating flame front produces a pressure wave or a rarefaction wave. It is seen that after 0.6ms, $\Delta U_+ = U_+(y=0.4) - U_+(y=0)$ becomes positive, and from this time central parts of the flame propagate with smaller velocities compared with the local velocities of the flame front, which are closer to the side walls, at $0 < y \leq 0.4\,\text{cm}$.

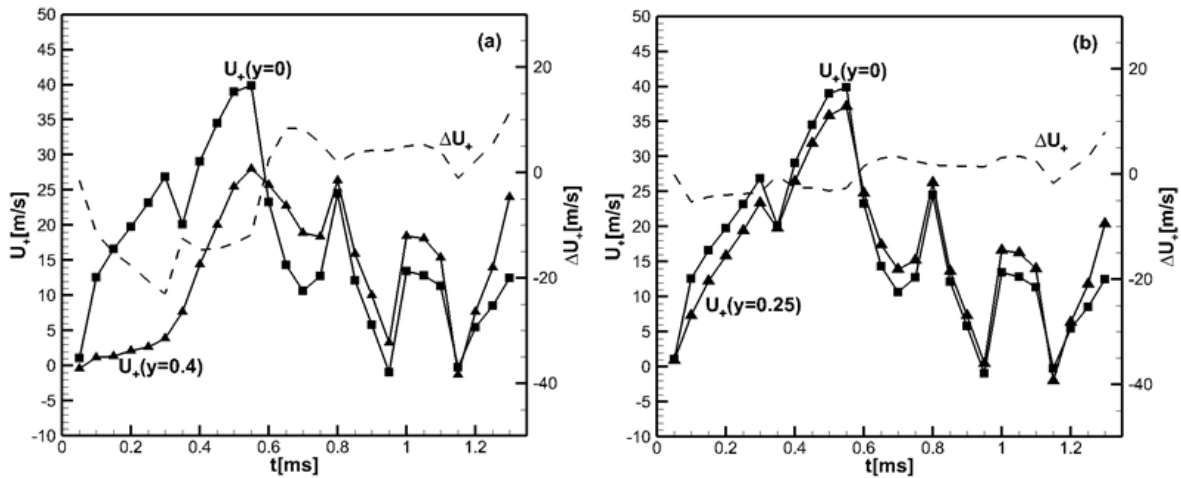

**Fig. 4 (a, b).** Velocities of the flow at 1 mm ahead of the flame front at $y=0;\ 0.4;\ 0.25\,\text{cm}$, and the difference $\Delta U_+$ between $U_+(y=0.4)$, $U_+(y=0.25)$ and $U_+(y=0)$.

It should be emphasized two important factors involved in the initiation of reverse flow in the unburnt gas by a rarefaction wave. The reverse flow behind the front of the rarefaction wave interacts with the existed previously "positive" flow, which leads to the decrease or reverse in the flow velocities and to the increase of the boundary layer thickness, which is maximal in the near-field flow ahead of the flame front, as it is shown in Figure 5. Obviously, the change in the flow speed is stronger in the bulk flow near the center line and less near the sidewall closer to the boundary layer. Since the local velocity of the flame front is the sum of the laminar flame velocity relative to the unburned mixture and the flow velocity in the near field ahead of the flame front, the change in the local velocity of the various parts of the flame front in the laboratory reference system will be as greater as closer the flame front point to the tube axis and minimum closer to the boundary layer.

Figure 5 shows the temporal evolution of the axial velocity profiles in the unburned gas, which creates conditions for the inversion of the flame front and the formation of a tulip flame. Due to the rarefaction wave the boundary layer thickness in the near-field zone ahead of the flame increases from less than 0.1 mm before 0.6ms to 0.1 cm and after 0.7ms the axial velocity profile in the unburned flow acquires a form of inverted tulip. After 1.35ms, the flame surface area increases and this process terminated.

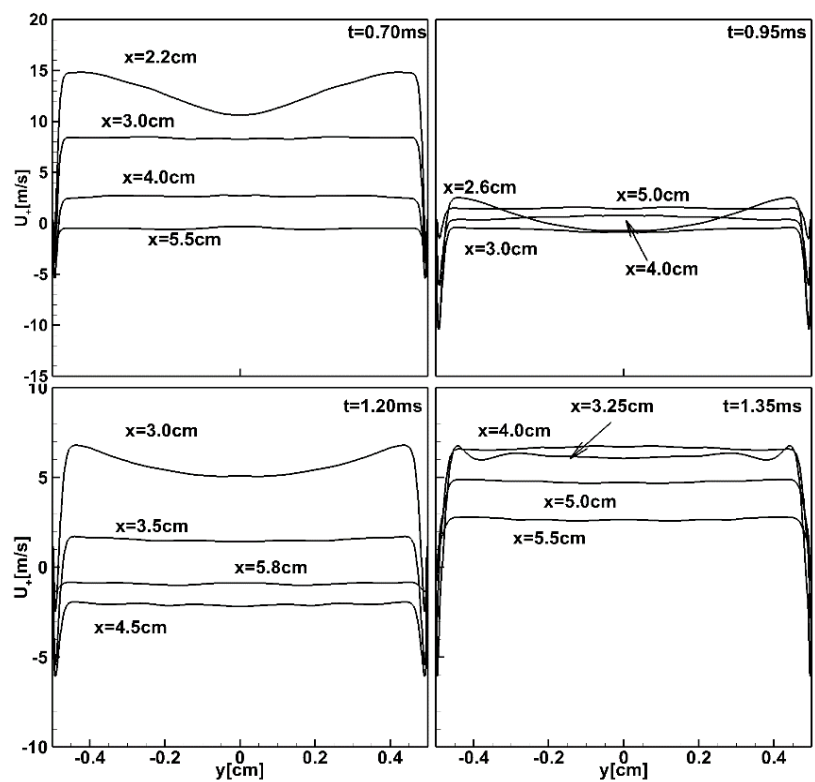

**Fig. 5.** Velocity profiles in the flow ahead of the flame in different cross sections of the tube.

It is seen in Figure 1 that the stage of flame front flattering, which begins approximately at 0.75 ms, is accompanied by the formation of two vortices that appeared in the burnt gases near the sidewalls. According to the theoretical model by Matalon and Metzener [24, 25], vortices play an important role in the formation of a tulip flame. The role of a pair of vortices, created by the highly curved parts of the flame near the sidewalls, apparently due to the baroclinic effect, have been discussed by many authors e.g. by Dunn-Rankin et al. [10, 11, 12]. The formation of vortexes was first mentioned by Dunn-Rankin and Sawyer [11]. Metzener and Matalon [24, 25] suggested a seemingly plausible scenario that: "The highly curved segments of the flame near the walls generate a pair of intense vortices in the burned gas that advects the flame into the tulip shape." It is known [52] that the baroclinic effect leads to the generation of vorticity when a curved interface between two matters of different densities is exposed to a pressure gradient. The baroclinic displacement is determined by the inviscid two-dimensional vorticity equation [52]

$$\frac{D\omega}{dt} = \frac{1}{\rho^2}[\nabla\rho \times \nabla P]. \tag{19}$$

When the flame front begins to flatten and the curvature of the flame front near the side walls becomes maximum, conditions become favorable for baroclinic vortex formation. This explains the generation of the first vortices in the combustion products near the side walls. We do not consider a possibility of vortices creation due to the Darrieus-Landau instability, but want to emphasized a pure hydrodynamic process that the boundary conditions require appearance of recirculation in the reversed flow of combustion products near the left closed end of the tube. These large scale vortical flow subsequently drifts in the burnt gas toward the flame front forming a stronger flow (the stream lines focusing) near the center line of the tube, as it is seen in Figure 1 at 0.85ms and 0.95ms. We can conclude that the physical process which explains and completely describes the formation of the tulip-shaped flame is the refraction waves generated by the flame during the deceleration stage. The vortical motion and the tulip/stagnation (saddle) point structure [28] are features of the flow, which was formed during the inversion of the flame front due to rarefaction waves generated by the decelerating flame.

These features are characteristics of the flow created and caused by refraction waves, but they are not the physical processes that cause the formation of a tulip flame.

The negative acceleration of the flame in general can create conditions for the Rayleigh–Taylor (RT) instability of the flame front. The effect of viscosity on the growth rate at high wavenumbers $k = 2\pi/\lambda$ is determined by equation [53]

$$\sigma_{RT} = \sqrt{kaA_t + k^4\bar{v}^2} - \bar{v}k^2, \tag{20}$$

where $A_t = (\rho_u - \rho_b)/(\rho_u + \rho_b) \approx 1$ is the Atwood number, and $\bar{v} = (\mu_1 + \mu_2)/(\rho_1 + \rho_2)$ is the average kinematic viscosity. The wavelength corresponding to the highest growth rate is determined by the expression [53]

$$\lambda_{max} = 4\pi\left(\frac{\bar{v}^2}{aA_t}\right)^{1/3}, \tag{21}$$

From this expression it is seen that the wavelength of the fastest growing mode is about 0.2cm. Figure 6 shows the calculated accelerations at various points $y = 0$, 0.25 and 0.4 cm at the flame front.

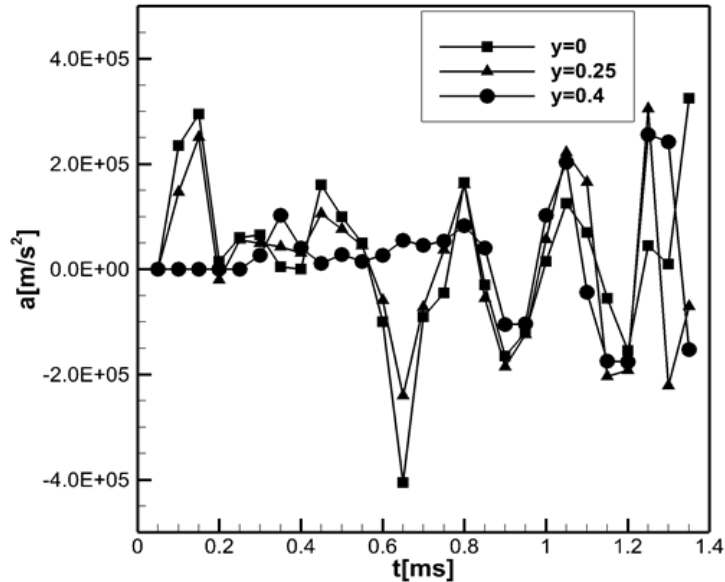

**Fig. 6.** Calculated accelerations at various points at the flame front during the tulip flame formation.

It can be seen that even for the greatest acceleration $a \approx 4\cdot 10^7$ cm/s, the product of the greatest growth rate and half of the oscillation period $T_{osc}$ during which the RT instability can develop, is $\sigma T_{osc} \approx 0.85$. This means that time $T_{osc}$ is not sufficient for the Rayleigh–Taylor instability development. In addition, the acceleration changes sign and has a stabilizing effect during the second half of the period.

Figures 7(a, b) show pressure $P(x, y=0)$ and flow velocity $U_+(x, y=0)$ profiles along the center line and near the boundary layer $U_+(x, y=0.4cm)$, for selected times during the tulip flame formation. The arrows at the Figures 7 show the direction of waves propagation. To trace the temporal evolution of the pressure gradient, additional pressure profiles are shown shortly before and shortly after the time shown in the upper right corner of each slide.

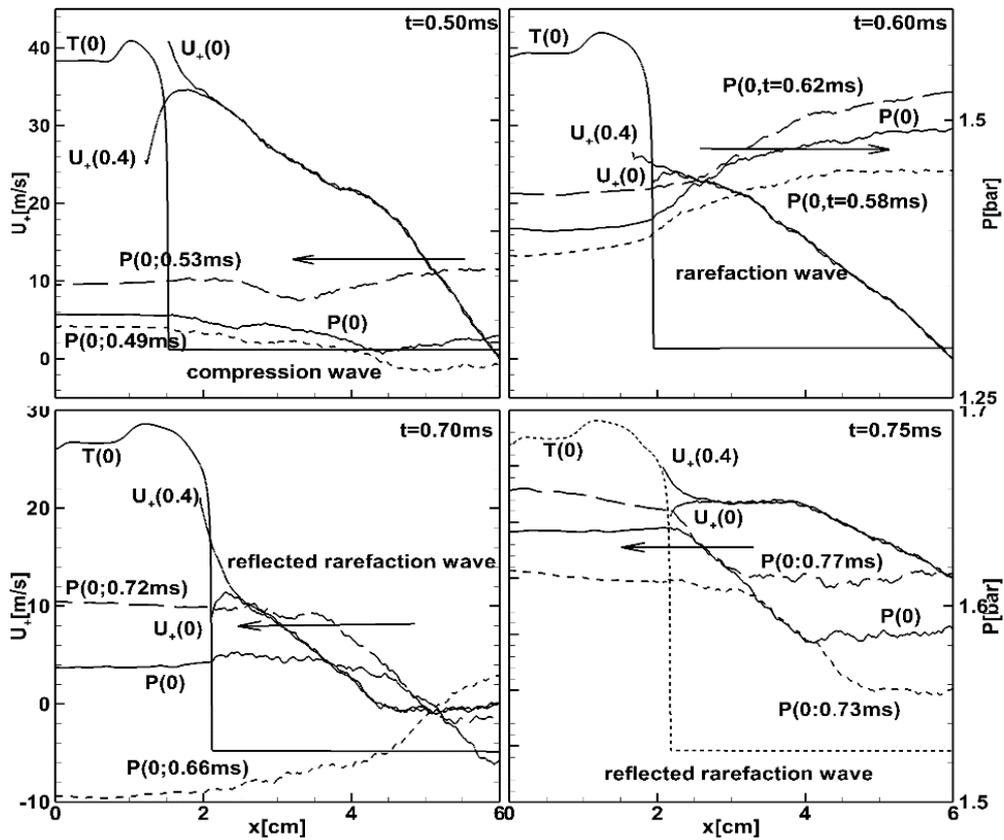

**Fig. 7(a)** Sequences of pressure and velocity distribution along the center line, and the flow velocity profiles $U_+(x, y=0)$ and $U_+(x, y=0.4)$ before the flame front flattering.

It is seen that at the beginning ($t < 0.6\,ms$), the flow velocity at the center line, $U_+(x, y=0)$ in the near-field unreacted gas mixture ahead of the flame front was larger than the velocity $U_+(x, y=0.4)$ near the sidewall. This results also in the larger local speed of the flame front along and near the central line compared with the flame front velocities near the wall. After 0.6ms the rarefaction wave drives the reverse flow to the left in the unreacted gas, leading to the increased thickness of the boundary layer in the near-field unburned regions (see Figure 5). The reverse flow in the burned gas is caused by the reverse flow from the unburned gas and it is supported by the built-up of the reverse (positive) pressure gradient at 0.7ms in Fig. 7(a).

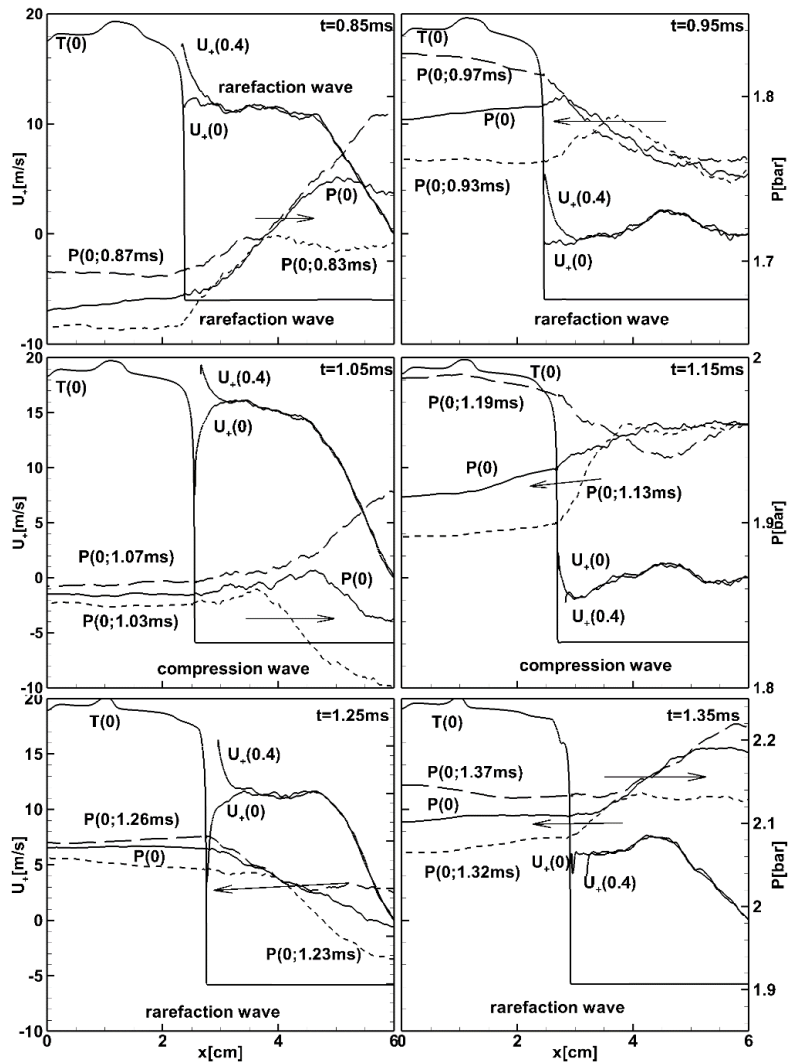

**Fig. 7(b)** Sequences of pressure and velocity distribution along the center line, and the flow velocity profiles $U_+(x, y=0)$ and $U_+(x, y=0.4)$ during the tulip flame formation.

Figure 7(b) shows time evolution of the pressure $P(x, y=0)$ and flow velocity $U_+(x, y=0)$ profiles along the center line and near the boundary layer $U_+(x, y=0.4cm)$ during the later stage corresponding to the flame front inversion and the tulip flame formation. It can be seen that after 0.85 ms the positive pressure gradient is almost "permanently" established, although it is sometimes smoothed by the reflected pressure waves. At the same time the flow velocity close to the boundary layer $U_+(x, y = 0.4 cm)$ in the near-field area of the unburnt flow exceeds the flow velocity at the center line flow $U_+(x, y = 0)$ all the way up to the tulip-shaped flame formation.

### 4.2 Effect of aspect ratio L/D =18

The velocity of the flame tip at the center line $y = 0$ computed for the flames propagating in tubes of various aspect ratios $L/D = 6, 12, 18$ with both closed ends and for the flame in the half-open tube (open right end) are shown in Figure 8. Arrows on the Figure 8 show some relatively minor decelerations of the flame tip velocity when the pressure waves reflected from the right closed end of the tube collide with the flame front for the first time at $t \approx 0.29\,\text{ms}, 0.62\,\text{ms}, 0.84\,\text{ms}$, for $L/D = 6, 12, 18$, respectively.

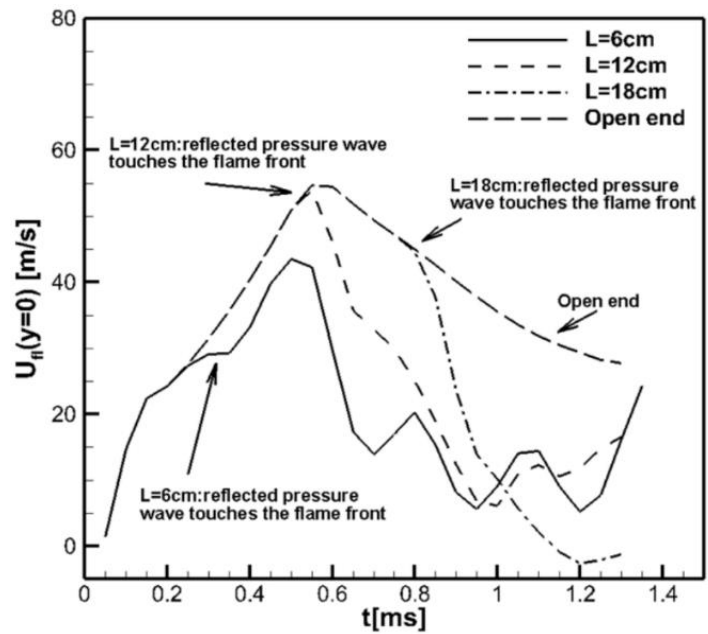

**Fig. 8.** The velocity of the flame tip along the center line $y = 0$, computed for tubes $L/D = 6, 12, 18$ and the half-open tube.

In a sense, the mechanism of the tulip flame formation is better seen for the flame propagating in the channel with larger aspect ratio and in a half-open tube. Figure 9 shows sequences of calculated schlieren images and streamlines at selected times during the development of a tulip flame for the channel $L/D = 18$.

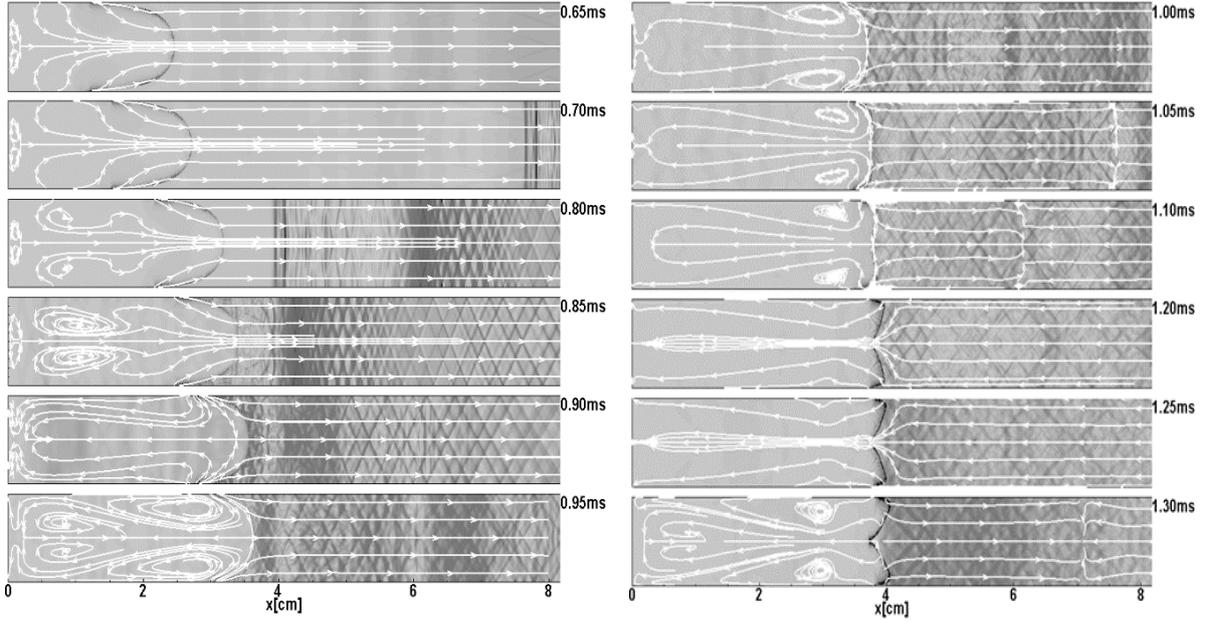

**Fig. 9** Schlieren images and stream lines for the premixed hydrogen/air flame propagating in the tube with both closed ends, aspect ratio $L/D = 18$.

Figure 10 shows the time evolution of local velocities of the flame front at the center line, $U_{fl}(y=0)$, near the side wall, $U_{fl}(y=0.4\text{cm})$, the combustion wave velocity $S_f$, and the flame front surface area $F_f$. It is seen that the flame surface begins to decrease at 0.55ms when the rear edge of the flame skirt touched the side wall. At 0.57 ms begins deceleration of the combustion wave and the flame tip. Compared to the shorter tube considered in Sec. 4.1, the rate of deceleration in this case is less, as was explained earlier. Therefore, the rarefaction wave is weaker and the flame front begins to flatten later than for a short tube, after 0.95ms. The deceleration becomes stronger with the stronger decrease of the flame surface after 0.95ms. During this time the combustion wave speed and the local speed of the flame front at the center line, $U_{fl}(y=0)$ decrease significantly, while the local speed of the flame front near the boundary layer, $U_{fl}(y=0.4\text{cm})$ increases. After 1.0ms the local flame front velocity at $y=0.4$ cm exceeds the velocity of the flame tip at $y=0$ and from that time the inversion of the flame front and the tulip flame begin to develop.

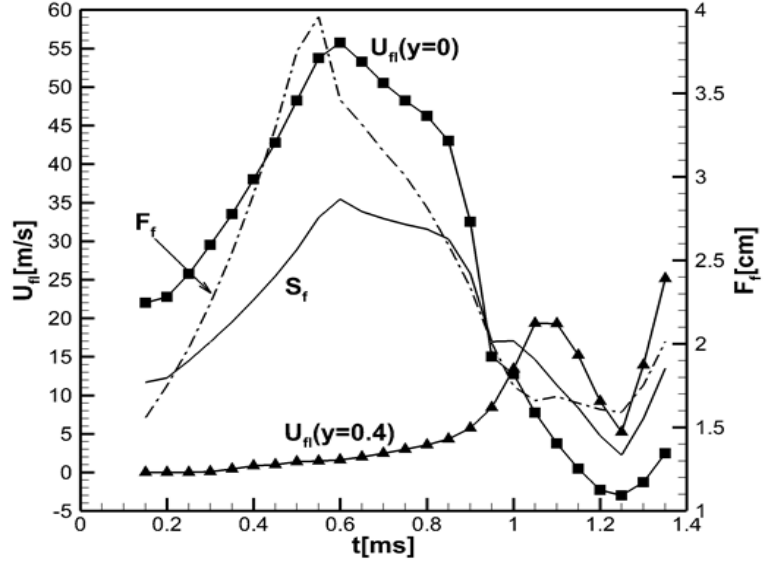

**Fig. 10**. Evolution of combustion wave velocity $S_f$, the flame surface area $F_f$, and local velocities of the flame front at $y=0$ and at $y=0.4$ cm for the tube $L=18$ cm.

In general, the entire scenario of the flame front inversion and the tulip-shaped formation is qualitatively similar to that considered in the previous section for $L/D=6$. The difference is that at greater aspect ratio there are fewer collisions of the flame front with pressure waves reflected from the right end of the tube, which makes clearer the decisive role of the rarefaction wave in the process of the flame front inversion from a convex to a concave shape. Since the transition time from the accelerating to the decelerating stage is determined by the decrease of the flame surface area due to collapse of the lateral part of the flame skirt touching the side walls, it depends mainly on the tube width and, for the same size of the ignition hot spot, it is almost the same for $L=6$ cm, $12$ cm, $18$ cm (see Fig. 8). It is slightly smaller for a short tube $L=6$ cm compared with longer tubes $L=12$ cm, $18$ cm because the flame front in the short tube collided with the reflected pressure wave, which slows down the flame before the transition to decelerating stage.

Figure 11(a, b) shows the evolution of the flow velocity at 1mm ahead of the flame front and the difference $\Delta U_+ = U_+(y=0) - U_+(y=0.4)$ between the velocities at the center line, and closer to the boundary layer at $y=0.4\,cm$ and $y=0.25\,cm$. The values of $\Delta U_+$ for all $0 < y \leq 0.4\,\text{cm}$ become positive after 0.95ms, and continue to increase up to the tulip-shaped flame formation.

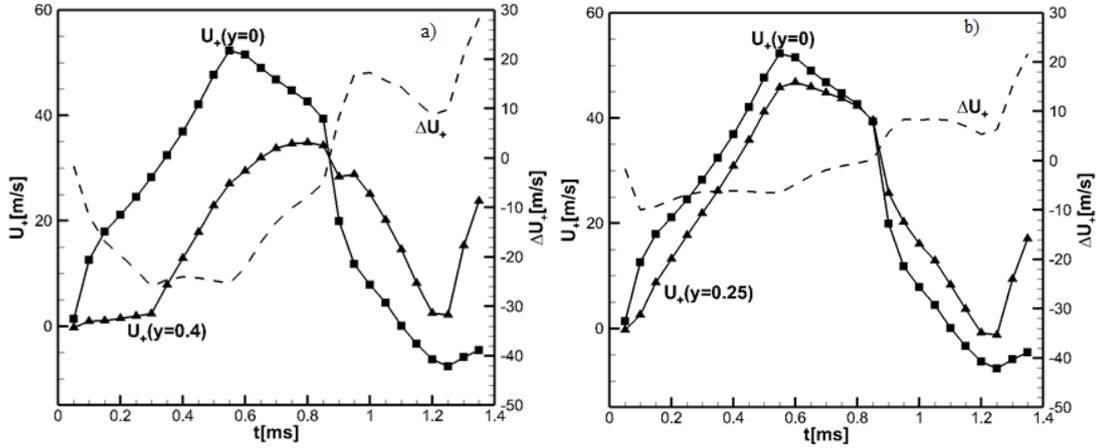

**Fig. 11(a, b)**. Velocities of the flow at 1 mm ahead of the flame front at $y = 0,\ 0.4,\ 0.25\,\text{cm}$ and the difference $\Delta U_+$ between $U_+(y=0.4), U_+(y=0.25)$ and $U_+(y=0)$; $L = 18\,\text{cm}$.

Figure 11(a, b) shows the evolution of the flow velocity at 1mm ahead of the flame front and the difference $\Delta U_+ = U_+(y=0) - U_+(y=0.4)$ between the velocities at the center line, and close or near the boundary layer, at $y = 0.4\,\text{cm}$ and $y = 0.25\,\text{cm}$. The values of $\Delta U_+$ for all $0 < y \leq 0.4\,\text{cm}$ become positive for $t > 0.95\,\text{ms}$, and continue to increase in absolute value during the flame front inversion. The difference between $U_+(y=0.4\,\text{cm})$ and $U_+(y=0)$ reaches maximum, when the rarefaction wave reverse the flow of the unburned gas ahead of the flame as this is seen in Fig. 9 for $t \geq 1.2\,\text{ms}$. As it was said earlier, in this case, the streamlines in the flow ahead of the flame front appear as "concentrated" or "converging" towards the center line $y = 0$ without appearance of vortexes in the unburned and in the burned flows. Figure 12 shows the change in the velocity profiles in the near-field unburned flow ahead of the flame and the increase in the thickness of the boundary layer caused by the interaction of the rarefaction wave with the previously formed positive flow created during the flame acceleration. It can be seen that as a result of the rarefaction wave in the flow ahead of the flame, the thickness of the boundary layer in the unburned gas flow increases significantly, and the velocity profile in the unburned gas flow takes the form of a reversed tulip, thereby turning the flame front into a tulip shape.

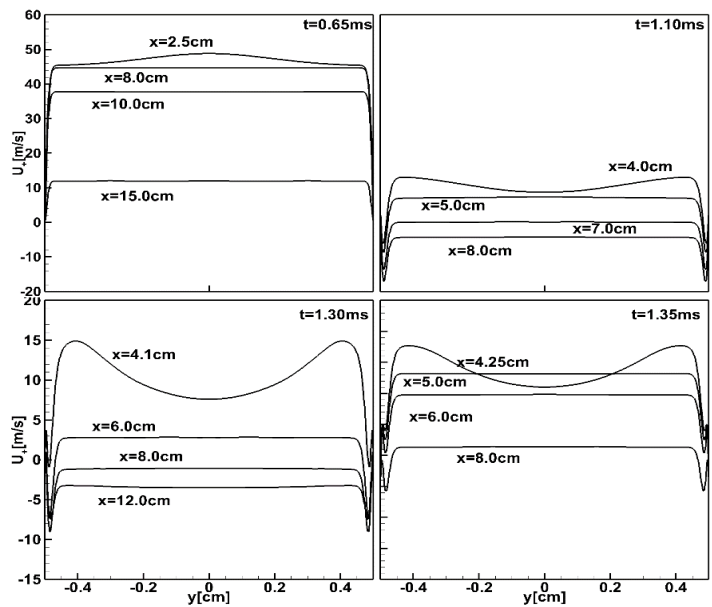

**Fig. 12.** Axial velocity profiles in the flow ahead of the flame front at selected times of the tulip flame formation.

Figure 13(a, b) shows the sequences of pressure profiles $P(x, y = 0)$ and the flow velocity along the center line $U_+(x, y = 0)$ and near the boundary layer $U_+(x, y = 0.4cm)$ during the tulip flame development.

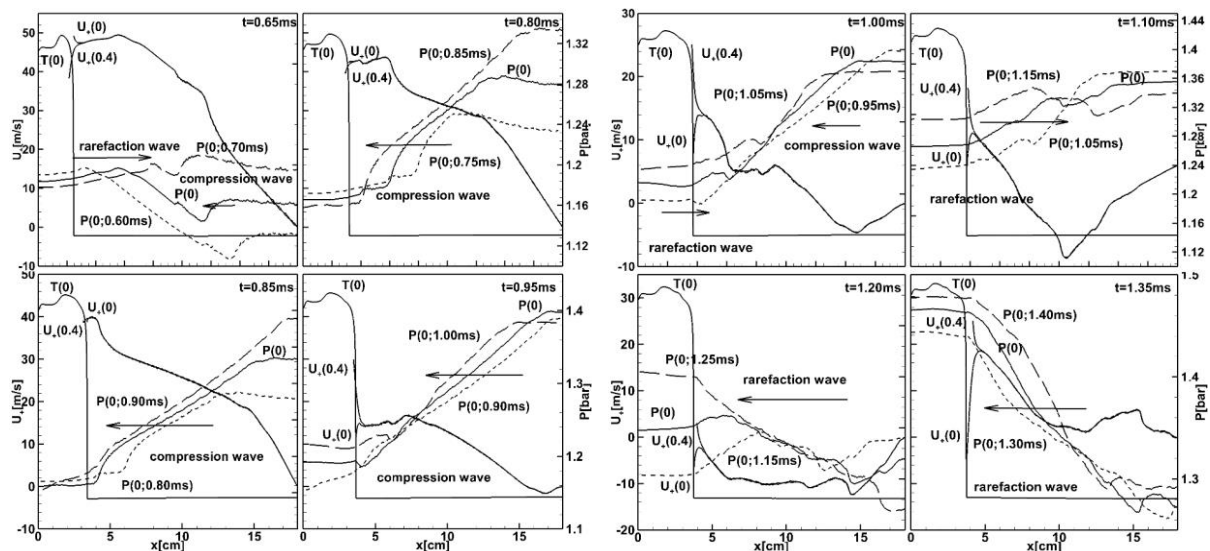

**Fig. 13(a, b).** Sequences of pressure and velocity distribution along the center line, and the flow velocity profiles $U_+(x, y = 0)$ and $U_+(x, y = 0.4)$; L=18cm.

It can be seen that before 0.65 ms, the unburned gas velocity at the center line still exceeds that at the wall, and the rarefaction wave is still too weak and the pressure gradient changes sign from positive to negative in the flow in front of the flame. After 0.95 ms, the rarefaction wave becomes sufficiently strong, and from this point the flow velocity near the boundary layer exceeds that at the center line, and the pressure gradient near the flame front remains positive until a tulip flame will be formed.

### 4.3 Tulip flame formation in a half-open tube

In the case of a half-open tube, when the flame is ignited at the left closed end and propagates to the right open end, the pressure waves leave the tube, there are no reflected pressure waves. As can be seen in Fig. 8, the acceleration stage in a half-open tube is almost the same as for the flame in closed tubes, but the rate of deceleration in the case of a half-open tube is noticeably lower than in tubes with both closed ends. In tubes, both ends of which are closed, the flame deceleration rate increases due to the increasing pressure. Therefore, in a half-open tube, only the rarefaction waves affect the dynamics of the flow ahead of the flame and, consequently, the formation of the tulip flame.

Figure 14 shows the time evolution of the flame surface area $F_f$, local velocities of the flame front along the central line and near the sidewall, at $y = 0.4\,\text{cm}$, and the speed of the combustion wave $S_f$, for the half-open tube.

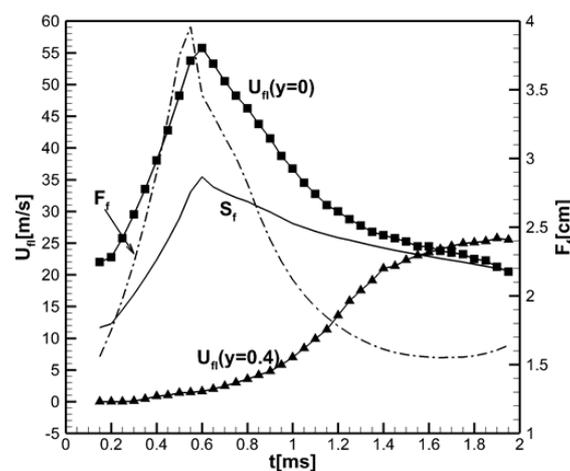

**Fig. 14.** Calculated time evolution of combustion wave velocity $S_f$, the flame surface area $F_f$ (dashed-dotted) and local velocities of the flame front at $y = 0$ and near the sidewall at $y = 0.4\,\text{cm}$ for the half-open tube.

It can be seen that the flame deceleration stage in a half-open tube lasts almost twice longer than in a tube with both ends closed. The inversion of the flame front and the formation of a tulip-shaped flame also last about two times longer compared to tubes with both ends closed, and the decrease in the flame surface area is monotonous and smoother.

Figure 15 (a, b) shows the sequences of calculated Schlieren images and streamlines during the development of a tulip flame for the flame propagating in the half-open tube.

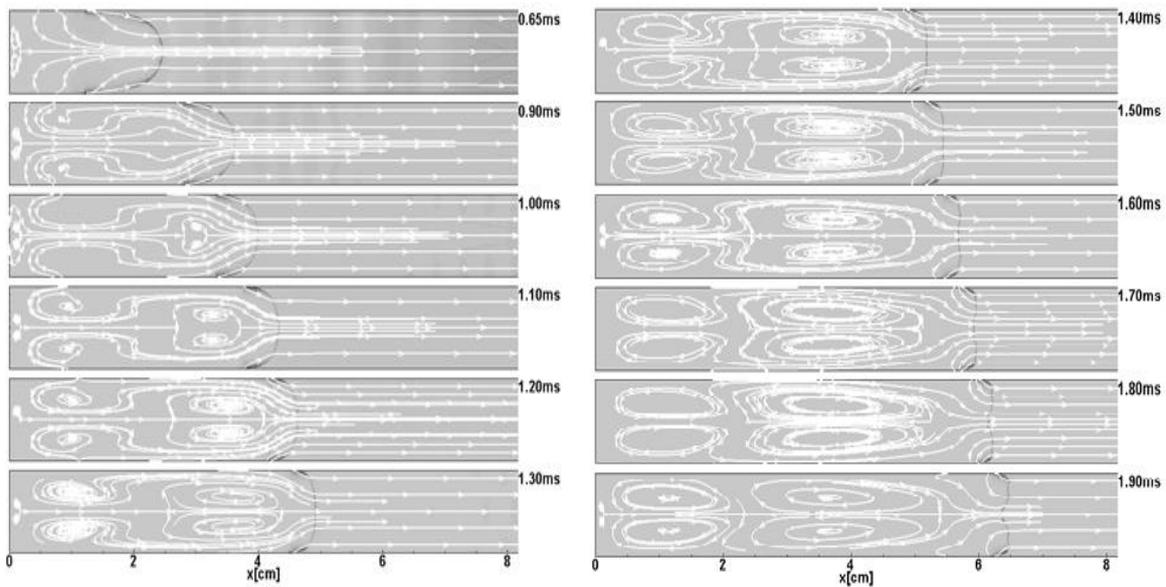

**Fig. 15(a, b)**. Sequences of computed Schlieren images and the stream lines for the premixed hydrogen–air flame propagating to the right open end.

Figure 16 shows the flow velocities in the unburned gas at 1 mm ahead of the flame front along the center line, at two points $y = 0.25cm$ and $y = 0.4cm$, and the differences $\Delta U_+ = U_+(y = 0.25cm) - U_+(0)$ and $\Delta U_+ = U_+(y = 0.4cm) - U_+(0)$.

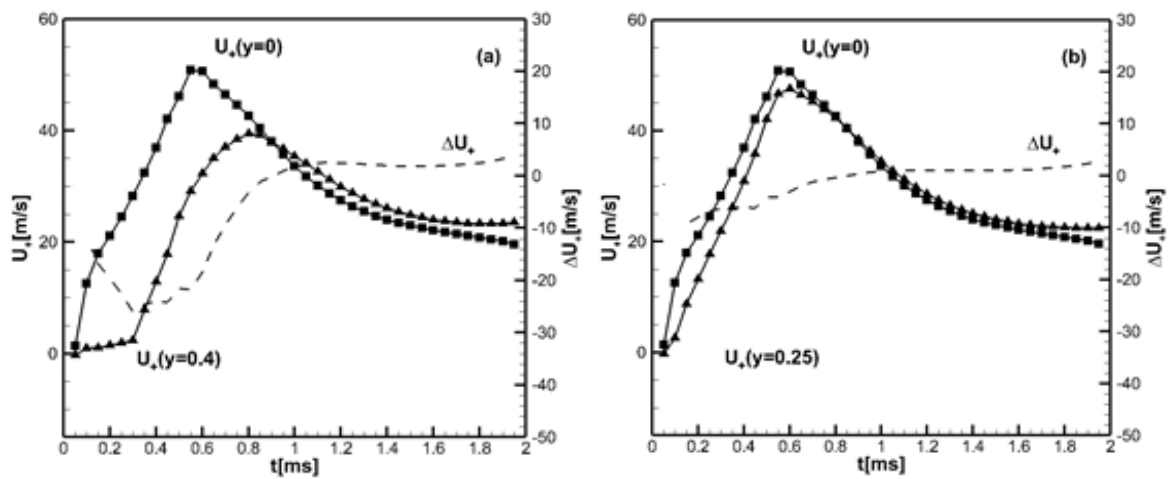

**Fig. 16(a, b).** Velocities of the flow at 1 mm ahead of the flame front at $y = 0, 0.25\ 0.4\,cm$ and the difference $\Delta U_+$ between $U_+(y = 0.4cm)$, $U_+(y = 0.25cm)$ and $U_+(y = 0)$; half open tube.

It can be seen that after 1 ms, the flow velocities in the unburned gas closer to the side walls exceed the velocity at the center line, and the farther from the center line, the greater the difference $\Delta U_+$. This trend continues all the way to the edge of the boundary layer, where the velocity drops to zero at the side wall of the tube. Accordingly, the local velocity of the flame front $U_{fl} = U_f + U_+$ is minimal at the central line and gradually increases toward the boundary layer, where it becomes maximum and then drops to zero inside the boundary layer.

Figure 17 shows the velocity profiles in the unburned flow in various cross sections of the tube ahead of the flame. It can be seen that the boundary layer thickness gradually increases, and when the flame front inverts to form a tulip-shaped flame, the boundary layer thickness increases to about 1 mm, from less than 0.1 mm at 0.65 ms. In fact, the location of the tips of the tulip petals is mainly determined by the thickness of the boundary layer, the laminar velocity of the flame, and the width of the tube.

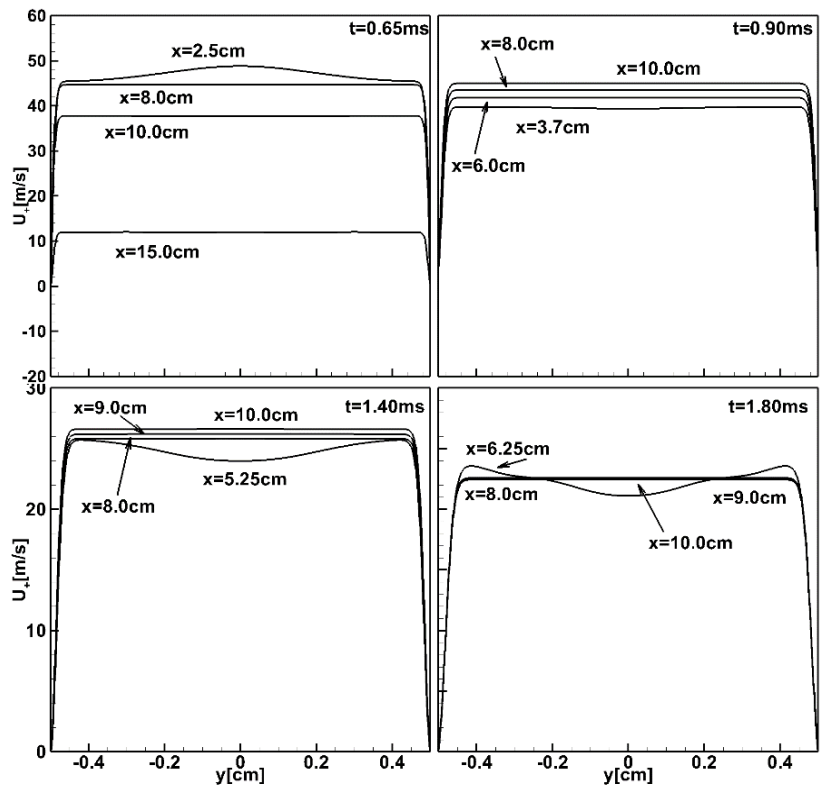

**Fig. 17.** Axial velocity profiles in the flow ahead of the flame in various cross sections of the half-open tube.

Figure 18 shows the dynamics of pressure profiles and velocities at the center line and close to the boundary layer in the flow ahead of the flame during the formation of the tulip flame. It can be seen that a negative pressure gradient exists only at the stage of flame acceleration, up to 0.65 ms. After 0.65 ms, when the flame surface begins to decrease and the decelerating flame generates rarefaction waves, a reverse (positive) pressure gradient is formed, which persists until the tulip flame is formed. Since the flame deceleration is weaker in the case of a half-open tube, the rarefaction waves created by the decelerating flame are also weaker than in the case of tubes with both ends closed. Therefore, in a half-open tube, rarefaction waves can reverse the pressure gradient and create a positive pressure gradient, but creating a reverse flow in front of the flame is more difficult.

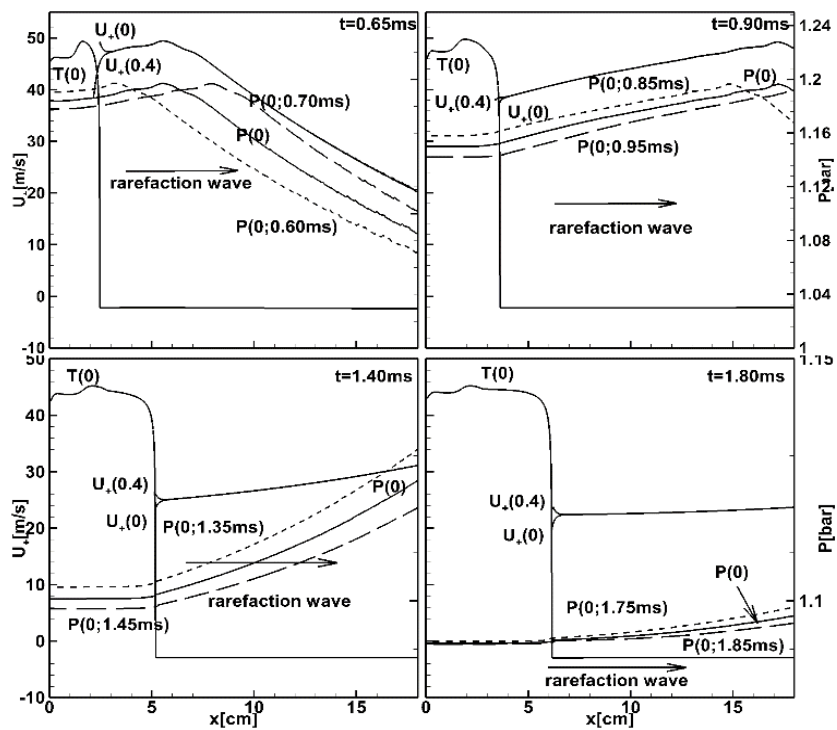

**Fig. 18**. Pressure and the flow velocity profiles ahead of the flame during the development of the tulip flame for half open tube.

## 5. 3D simulations of tulip flame formation in a tube L/D=6

The flame dynamics and the tulip flame formation in a three-dimensional tube is qualitatively similar to that in a two-dimensional tube. The most significant difference is the higher rate of flame acceleration and deceleration, so that the maximum flame speed achieved in the acceleration stage is almost two times higher in 3D than in the 2D case, and the duration of the acceleration and deceleration stages is almost 1.5 times shorter. The greater acceleration of a three-dimensional flame in a tube with no-slip walls compared to a two-dimensional flame is due to the fact that the acceleration (deceleration) of the flame is determined by an increase (decrease) in the surface area of the flame, which is a line in the 2D case and a surface in the 3D case. This phenomenon was discussed in [53, 54] in connection with modeling the transition from deflagration to detonation. Figure 19 shows the change in time of the flame surface area $F_f$, the velocities of the flame front $U_{fl}(y=0)$ and $U_{fl}(y=0.3\,\text{cm})$, and the combustion wave velocity $S_f$, for a flame propagating in a tube of length $L=6\,cm$ and cross section $D \times D = 1\,\text{cm}^2$. It can be seen that qualitatively this figure is similar to Fig. 2 obtained for the 2D case. In the 2D case the flame surface area begins to decrease at 0.5ms, and the decelerating stage begins at 0.55ms (Fig. 2), while in the 3D case for the tube of the same aspect ratio, these times are 0.33ms and 0.38ms. In particular, the pressure wave reflected from the right closed end does not have time to collide with the flame during the deceleration stage, and there are no large oscillations of the velocities caused by the reflected pressure waves.

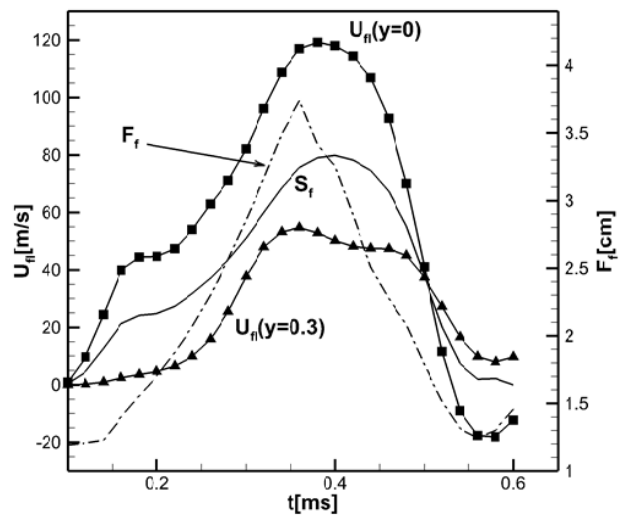

**Fig. 19.** Time evolution of the flame surface area $F_f$ (dashed-dotted), local velocities of the flame front at $y=0$ and at $y=0.3\,cm$, and combustion wave velocity $S_f$, for the tube with both closed ends, $L/D=6$.

Figure 20 (a, b) shows the flow velocities at the cross section $(x, y, z = 0)$ at 1mm ahead of the flame front at different distances from the central line $y = z = 0$, at $y = 0.15$ cm and $y = 0.3$ cm, and $\Delta U_+ = U_+(y = 0.15 cm) - U_+(y = 0)$ and $\Delta U_+ = U_+(y = 0.3 cm) - U_+(y = 0)$. It is seen that the local velocities of the flame front far away from the central line begin to exceed the flame velocity along the tube axis ($y = z = 0$) after about 0.42ms. It is seen that the local velocities of the flame front far away from the central line begin to exceed the flame velocity along the tube axis ($y = z = 0$) after about 0.42ms. This difference increases in the direction from the tube axis and reaches a maximum at 0.3 cm from the tube axis, resulting in an inverted tulip-shaped profile of the axial velocity. Eventually, this leads to the formation of a tulip-shaped flame, with the tips of the petals forming at a distance of 0.3 cm from the axis of the tube. After 0.5 ms, the unburned flow velocity at the walls exceeds the flow velocity along the tube axis, the convex flame front begins to flatten, and then a tulip-shaped flame develops very quickly.

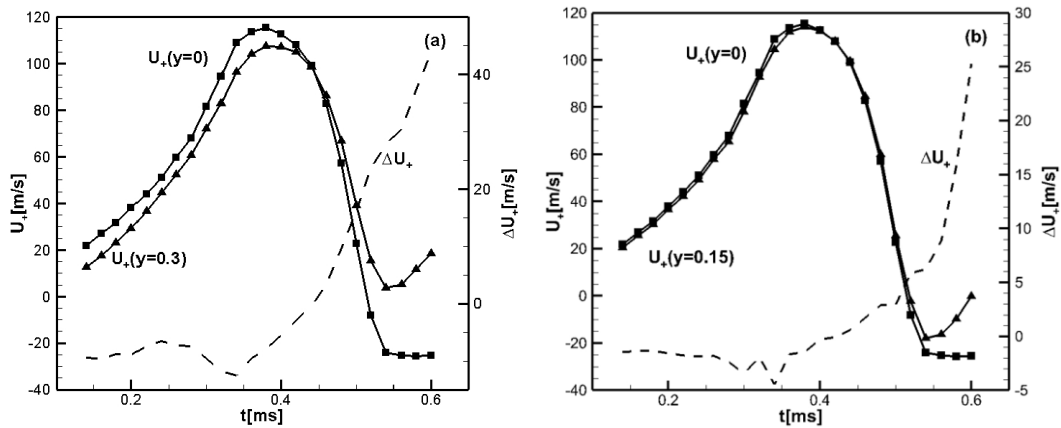

**Fig. 20 (a, b)**. The flow velocities at 1mm ahead of the flame front at the cross section $(x, y, z = 0)$, at $y = 0$, $y = 0.15$ cm, and $y = 0.3$ cm; and $\Delta U_+ = U_+(y = 0.15 cm) - U_+(y = 0)$ and $\Delta U_+ = U_+(y = 0.3 cm) - U_+(y = 0)$.

Figure 21 shows a sequence of numerical Schlieren images with the stream lines calculated for a flame propagating in the 3D channel, $L/D = 6$, $D \times D = 1 \text{cm}^2$. It can be seen that the convex flame front begins to flatten at 0.5 ms, which is accompanied by the formation of a pair of vortices near the flame front and the reverse flow of the burned gas along the side walls. The interaction of the reverse flow near the side walls with the remaining "positive" flow near the tube axis creates a large-scale recirculation - a pair of vortices near the left closed end of the tube.

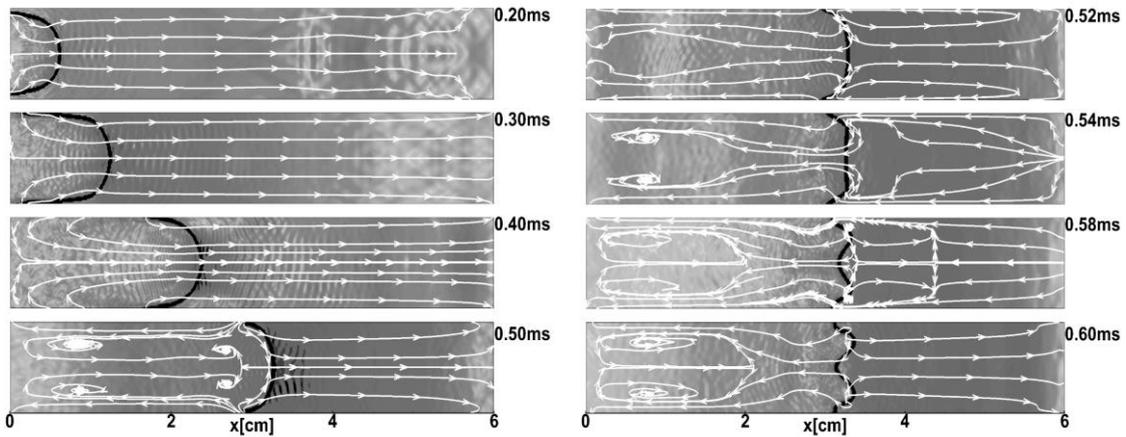

**Fig. 21 (a, b).** Schlieren images and stream lines; premixed hydrogen/air flame propagating in the 3D tube with both closed ends $L = 6\,\text{cm}$, cross section $1 \times 1\,\text{cm}^2$; shown is cross section $(x, y, z = 0)$.

Figure 22 shows the velocity profiles in the unburned mixture ahead of the flame in various cross sections of the tube. It is seen that the rarefaction wave causes the increase in the thickness of the boundary layer in the unburned flow near the flame front. After 0.5 ms, the velocity profile in the near-field zone ahead of the flame has the appearance of an inverted tulip. Therefore, the local velocity of the flame front, $U_{fl}(y) = U_f + U_+(y)$, becomes minimal at the tube axis, $y = 0$, gradually increases toward the side walls to a maximum near the boundary layer, at $y = 0.3\,\text{cm}$, drops to zero at the sidewall, at $y = 0.5\,\text{cm}$, and the tulip-shaped flame is formed.

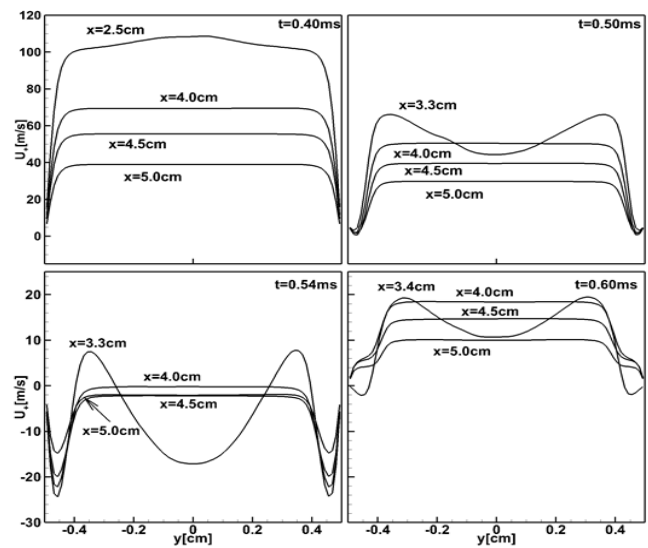

**Fig. 22.** 3D LES simulations: velocity profiles in the flow ahead of the flame in different cross sections of the tube.

Figure 23 shows the time evolution of the pressure gradient across the flame front and the velocities in the unburned flow at the tube axis ( $x, y = 0, z = 0$ ) and near the side walls ( $x, y = 0.3$ cm, 0). It is seen that the initially negative pressure gradient during the accelerating stage at $t < 0.4$ ms, turns into the positive pressure gradient at 0.42ms, when decelerating flame generated a rarefaction wave, and it remains positive up to 0.58ms, when the flame surface area begins to increase and the formed tulip flame begins to accelerate.

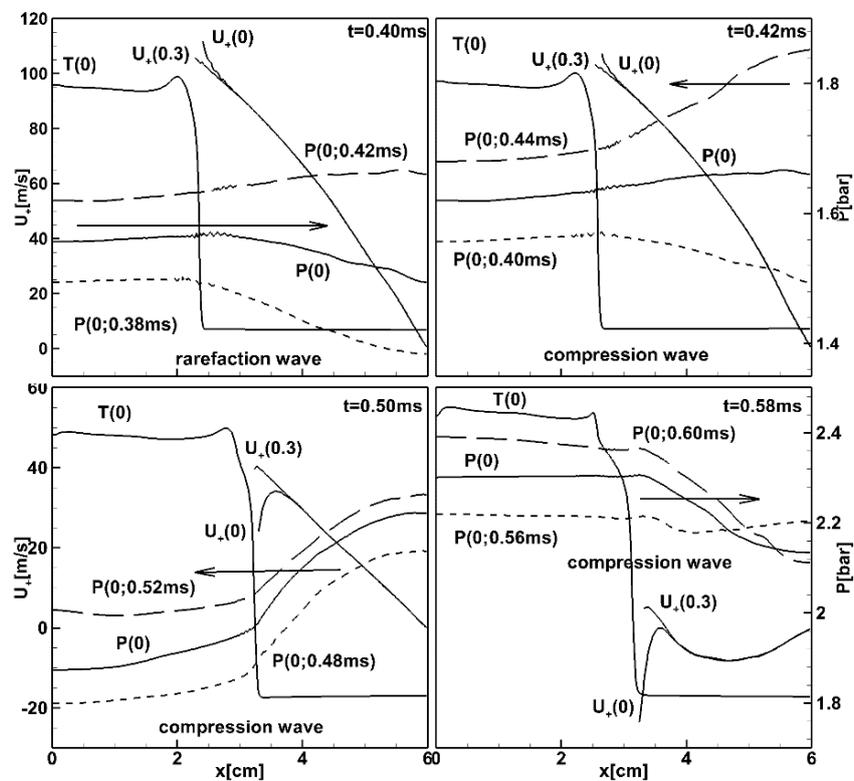

**Fig. 23.** Sequences of pressure profiles along the center line and the flow velocities ahead of the flame at $y = 0$ and $y = 0.3$ cm; 3D tube $L = 6$ cm, cross section $1 \times 1$ cm$^2$.

## 6. Summary and Conclusions

In this paper we presented results of 2D direct numerical simulations and 3D Large eddy simulations with a detailed chemical model describing propagation of a stoichiometric hydrogen-air flame and the formation of tulip flames in tubes of various aspect ratios $L/D = 6, 12, 18$ with both ends closed and in the half-open tube. Numerical simulations show a close similarity between the numerical and experimental results available in the literature, indicating that the simulations correctly reproduce the dominant physical processes involved in the flame dynamics of tulip flame formation. It was found that the formation of a tulip flame is a pure hydrodynamic process, which does not involve any of the intrinsic instabilities of the flame front in agreement with the conclusion reached by Ponizy et al. [27] based on the experimental study of the tulip flame formation in a stoichiometric propane-air flame propagating in a cylindrical tube closed at both ends.

It is shown that the rarefaction waves produced by the flame during deceleration stage when the surface area of the flame begins to decrease due to the collapse of the lateral parts of the flame skirt on the side walls of the tube, play a key role in the inversion of the flame front and the formation the tulip flame. The flattening of the flame front is accompanied by the creation of the first pair of vortices near the side walls and the reversal of the burned gas flow. Later-formed large scale circulations in the reversed flow of the burned gas are required by the continuity and boundary conditions. However, neither the vortical motion nor the intrinsic flame front instabilities are involved in the formation of the tulip flame.

The interaction of rarefaction waves with the unburned gas ("positive") flow, established at the acceleration stage, leads to the reduction and reversal of flow velocity, as well as to the decrease of the flow velocity in the near-field zone ahead of the flame front, and to the increase in the thickness of the boundary layer. This leads to the formation of the axial velocity profile in the near-field region ahead of the flame in the form of an inverted tulip. Since the velocity of every parts of the flame front in the laboratory reference frame is determined by the local velocity of the unburned gas in the near-field zone just ahead of this part of the flame front, such the axial velocity profile leads to the flame front inversion and the tulip flame formation. From the above-described mechanism of the tulip flame formation it is obvious that the tips of the formed flame tulip petals are at a distance equal to the thickness of the boundary layer in the near-field region ahead of the flame, which is confirmed by simulations.

The formation of a reverse (adverse) pressure gradient, which was discussed by Xiao et al. [31] is a natural effect of the rarefaction waves produced by a decelerating flame, since the pressure gradient in the rarefaction wave is positive unlike the negative pressure gradient created by pressure (compression) waves generated by the flame during accelerating stage. The front of a rarefaction wave propagates forward with the speed of sound while the gas flow behind the front of the rarefaction wave propagates in the opposite direction, therefore the pressure gradient in the rarefaction wave is positive.

The three-dimensional LES modeling of the tulip flame formation is qualitatively similar to the results of the 2D direct numerical simulation, demonstrating the same mechanism of the tulip flame formation. The main difference between 2D and 3D cases is that the characteristic times in the 3D case is almost twice less than that in the 2D case. This means that one can expect only a qualitative, but not quantitative agreement of the 2D simulations with the experimental study of the tulip flame.

**Acknowledgements**

This work was supported by National Natural Science Foundation of China under grant 11732003 (C.Q and C.W.). The research of Nordic Institute for Theoretical Physics (NORDITA) is partially supported by Nordforsk. No any specific grant was received (M.L.)

**Disclosure statement**

No potential conflict of interest was reported by the authors.